\documentclass[11pt]{article}
\usepackage{amssymb,graphicx}

\newcommand{\edo}{\end{document}}
\newcommand{\extraref}[1]{}  

\newcommand{\R}{{\mathbb R}}  

\newcommand{\tc}{{\cal T}}

\newcommand{\KX}{K^X}  
\newcommand{\KY}{K^{\mathcal{Y}}}
\newcommand{\KU}{K^{\mathcal{U}}}

\newcommand{\kx}{k^X}  
\newcommand{\ky}{k^{\mathcal{Y}}}

\newcommand{\ip}[2]{\langle #1 , #2 \rangle}

\def\ben{\begin{enumerate}}
\def\een{\end{enumerate}}

\newtheorem{theorem}{Theorem}
\newtheorem{itlemma}{Lemma}[section] 
\newtheorem{itproposition}[itlemma]{Proposition}
\newtheorem{itcorollary}[itlemma]{Corollary}
\newtheorem{itremark}[itlemma]{Remark}
\newtheorem{itdefinition}[itlemma]{Definition}
\newtheorem{itexample}[itlemma]{Example}

\newenvironment{lemma}{\begin{itlemma}\rm}{\end{itlemma}} 
\newenvironment{remark}{\begin{itremark}\rm}{\end{itremark}} 
\newenvironment{corollary}{\begin{itcorollary}\rm}{\end{itcorollary}}
\newenvironment{proposition}{\begin{itproposition}\rm}{\end{itproposition}}
\newenvironment{definition}{\begin{itdefinition}\rm}{\end{itdefinition}}
\newenvironment{example}{\begin{itexample}\rm}{\end{itexample}}
\newenvironment{proof}{\noindent {\em Proof}.\
}{\hspace*{\fill}$\halmos$\medskip}

\newcommand{\text}[1]{\hbox{\rm \ #1\ \/}}
\newcommand{\be}[1]{\begin{equation}\label{#1}}
\newcommand{\ee}{\end{equation}}
\newcommand{\beqn}{\begin{eqnarray*}}
\newcommand{\eeqn}{\end{eqnarray*}}
\newcommand{\beq}{\begin{eqnarray}}
\newcommand{\eeq}{\end{eqnarray}}
\newcommand{\bl}[1]{\begin{lemma}\label{#1}}
\newcommand{\ble}[1]{\begin{lemmaex}\label{#1}}
\newcommand{\br}[1]{\begin{remark}\label{#1}}
\newcommand{\bt}[1]{\begin{theorem}\label{#1}}
\newcommand{\bd}[1]{\begin{definition}\label{#1}}
\newcommand{\bp}[1]{\begin{proposition}\label{#1}}
\newcommand{\bc}[1]{\begin{corollary}\label{#1}}
\newcommand{\bfact}[1]{\begin{fact}\label{#1}}
\newcommand{\ber}[1]{\begin{exercise}\label{#1}}
\newcommand{\bex}[1]{\begin{example}\label{#1}}
\newcommand{\bem}[1]{\begin{example}\label{#1}}  
\newcommand{\ec}{\mybox\end{corollary}}
\newcommand{\efact}{\mybox\end{fact}}
\newcommand{\eer}{\mybox\end{exercise}}
\newcommand{\eex}{\mybox\end{example}}
\newcommand{\eem}{\mybox\end{example}}
\newcommand{\el}{\mybox\end{lemma}}
\newcommand{\ele}{\mybox\end{lemmaex}}
\newcommand{\er}{\mybox\end{remark}}
\newcommand{\et}{\qed\end{theorem}}
\newcommand{\ed}{\mybox\end{definition}}
\newcommand{\ep}{\mybox\end{proposition}}
\newcommand{\epr}{\end{proof}}
\newcommand{\bpr}{\begin{proof}}

\newcommand{\ecs}{\end{corollary}}
\newcommand{\eers}{\end{exercise}}
\newcommand{\eexs}{\end{example}}
\newcommand{\eems}{\end{example}}
\newcommand{\els}{\end{lemma}}
\newcommand{\eles}{\end{lemmaex}}
\newcommand{\ers}{\end{remark}}
\newcommand{\ets}{\end{theorem}}
\newcommand{\eds}{\end{definition}}
\newcommand{\eps}{\end{proposition}}
\newcommand{\halmos}{\rule{1ex}{1.4ex}}
\newcommand{\qed}{\hfill \halmos} 
\newcommand{\mybox}{\hfill $\Box$} 

\newcommand{\banach}{{\mathbb B}}
\newcommand{\st}{\; | \;}
\newcommand{\Inputset}{{\mathcal U}}
\newcommand{\Outputset}{{\mathcal Y}}
\newcommand{\edeq}{\vskip-0.7cm\mybox\end{definition}}

\def\bi{\begin{itemize}}
\def\ei{\end{itemize}}
\newcommand{\arrowschem}[2]{\raisebox{-1ex}%
	{$\stackrel{\stackrel{#1}{\longrightarrow}}%
	{\stackrel{\longleftarrow}{#2}}$}}
\newcommand{\fract}[2]{{#1}/{#2}}  

\begin{document}

\title{%
Multi-Stability in Monotone Input/Output Systems}
\author{David Angeli\\
Dip. Sistemi e Informatica \\
University of Florence,  50139 Firenze, Italy \\
{\tt angeli@dsi.unifi.it} \and
Eduardo D. Sontag%
\thanks{Supported in part by AFOSR Grant F49620-01-1-0063,
NIH Grants R01 GM46383 and P20 GM64375, and Aventis}
\\
Dept. of Mathematics\\
Rutgers University, NJ, USA\\
{\tt sontag@hilbert.rutgers.edu}}

\date{}
\maketitle 
\textbf{Abstract}  
This paper studies the emergence of multi-stability and hysteresis in those
systems that arise, under positive feedback, starting from monotone systems
with well-defined steady-state responses.
Such feedback configurations appear routinely in several fields of
application, and especially in biology.
Characterizations of global stability behavior are stated in terms of easily
checkable graphical conditions.


\section{Introduction}

Multi-stability and associated hysteresis effects form the basis of many models
in molecular biology, in areas such as cell differentiation, development,
and periodic behavior described by relaxation oscillations.
See for instance the classic work by Delbr\"uck~\cite{delbruck}, who suggested
in 1948 that multi-stability could explain cell differentiation, as well as
references in the current literature (e.g.,  
\cite{bhalla},
\cite{ferr},
\cite{ferrel},
\cite{nature},
\cite{hunding},
\cite{laurent},
\cite{pomerening},
\cite{thomas}).

One appealing class of systems in which to study this phenomenon is that of
{\em monotone systems with inputs and outputs}, a class of systems introduced
recently in~\cite{monotone1}, motivated by applications in molecular biology
modeling.
Monotone systems with inputs and outputs constitute a natural generalization
of classical (no inputs and outputs) monotone dynamical systems, which are
those for which flows preserve a suitable partial ordering on states.
The work reported here is grounded upon the rich and elegant theory of
monotone dynamical systems (see the textbook by Smith~\cite{smith} as well as
papers such as~\cite{Hirsch,Hirsch2} by Hirsch and \cite{smale} by Smale),
which provides results on generic convergence to equilibria, and, more
generally, on the precise characterization of omega limit sets.
One of the main difficulties in applying the theory of monotone dynamical
systems is that of determining the locations and number of steady states.
In this paper, we propose the idea of viewing more complicated systems as
positive feedback loops involving monotone systems with inputs and outputs
and well-defined steady state responses.  The feedback configuration may 
induce multiple steady states, and we show how the locations and stability of
them can be completely characterized using a simple planar graphical test.

We present the general theory, and illustrate the results by means of two
examples.  The first is a simple two-dimensional system; since such
systems can be analyzed using classical phase-plane techniques, the example
can be related to routine and elementary calculations, and thus the meaning of
our conditions is easy to understand.  The second example is of high order,
and arises in the study of cellular signaling cascades.  Further applications
are developed in the reference~\cite{angeli-ferrell}.

The organization of this paper is as follows.
First, we review the basic definitions regarding monotone systems and 
state our main results regarding positive feedbacks. 
Then we present and prove several graphical tests which are useful in
checking the properties required by our results.
After this, we provide proofs of the main theorems as well as a number of
needed technical facts concerning linear systems which arise when
linearizing general monotone systems along trajectories.
This is followed by two examples, as discussed above.
The paper closes with a discussion of hysteresis behavior, as well as a
subtle counterexample showing that monotonicity plays a crucial role and
cannot be dispensed with as an assumption.
Two appendixes contain proofs of some technical points.

\section{Basic Definitions}

We briefly review some of the main concepts and notations
from~\cite{monotone1}. 

By a {\em positivity cone} $K$ in a Euclidean space $\banach$
we mean a nonempty closed convex and pointed ($K\bigcap -K=\{0\}$) cone $K\subset \banach$.
In this paper, we assume that cones have nonempty interiors.
Associated to such a cone, one introduces a partial ordering:
$x_1 \succeq x_2$ (or ``$x_2 \preceq x_1$'') iff $x_1-x_2 \in K$.
Strict ordering is denoted by $x_1 \succ x_2$, meaning that 
$x_1 \succeq x_2$ and $x_1 \neq x_2$.
One also introduces a stricter ordering by the rule:
$x_1 \gg x_2 \Leftrightarrow x_1-x_2 \in \textrm{int}(K)$. 
A typical example is $\banach=\R^n$ with the ``NorthEast'' ordering given by
the first orthant: $K= \R_{\geq 0}^n$, in which case ``$x_1 \succeq x_2$'' means
that each coordinate of $x_1$ is bigger or equal than the corresponding
coordinate of $x_2$.  In this case, $x_1 \gg x_2$ means that every coordinate
of $x_1$ is strictly larger than the corresponding coordinate of $x_2$, in
contrast to ``$x_1 \succ x_2$'' which means only that some coordinate
is strictly larger.

State-spaces for monotone systems are by definition subsets $X$ of $\R^n$, for
a suitable $n$, and endowed with an order arising from a cone $\KX$
(or just ``$K$'' if clear from the context).
We assume always that $X$ is the closure of an open subset of $\R^n$.
Input sets $\Inputset$ are subsets of ordered spaces $\R^m$, and we write
$u_1 \succeq u_2$ whenever $u_1 - u_2 \in \KU$ where $\KU$ is the
corresponding positivity cone in $\R^m$,
for any pair of input values $u_1$ and $u_2 \in \Inputset$.
An ``input'' is a locally essentially bounded Lebesgue measurable function
$u(\cdot ):\R_{\geq 0}\rightarrow \Inputset$, and we write
$u_1  \succeq u_2$ provided that
$u_1 (t) \succeq u_2(t)$ for almost all $t\geq 0$.
Similarly, output sets $\Outputset$ will be assumed to be ordered as well.
To keep the notation simple and only when there is no risk of ambiguity,    
we use the same symbol for all orders.  

A (finite-dimensional continuous-time)
{\em system} in the sense of control theory (see e.g.\ \cite{mct})
\be{cds}
\dot x \,=\, f(x,u)\,,\quad y = h(x)
\ee
is specified by a state space $X$, an input set $\Inputset$, and an output set
$\Outputset$, 
where the map $f$ is defined on $\widetilde X\times \Inputset$, where $\widetilde X$ is some open
subset of $\R^n$ which contains $X$.
In general, one may assume that $f(x,u)$ is continuous in $(x,u)$ and locally
Lipschitz continuous in $x$ locally uniformly on $u$, but for simplicity in
this paper, we will assume that $f(x,u)$ is differentiable.  
In order to obtain a well-defined controlled dynamical system on $X$,
we will assume that the solution $x(t)=\phi (t,\xi ,u)$ (or just ``$x(t,\xi ,u)$'')
of $\dot x(t)=f(x(t),u(t))$ with initial condition $x(0)=\xi $ is defined for all
inputs $u(\cdot )$ and all times $t\geq 0$.
This means that solutions with initial states in $X$ must be defined for all
$t\geq 0$ (forward completeness) and that the set $X$ is forward invariant.
We say that the system is {\em monotone} if the following property holds, with
respect to the orders on states and inputs: 
\[ 
\xi _1 \succeq \xi _2 \quad \& \quad u_1 \succeq u_2 \quad\Rightarrow \quad x(t,\xi _1,u_1)
\succeq x(t,\xi _2,u_2) \quad \forall \, t \geq 0. 
\]
If $\textrm{int} (K) \neq \emptyset$, this is equivalent to asking:
\[ 
\xi _1 \gg \xi _2 \quad \& \quad u_1 \succeq u_2 \quad\Rightarrow \quad
x(t,\xi _1,u_1) \gg x(t,\xi _2,u_2)
\quad \forall \, t \geq 0 
\] 
(a set which is the closure of its interior is invariant iff its interior
is invariant, see~\cite{monotone1}).  
We also assume given a monotone ($x_1\succeq x_2$ $\Rightarrow $ $h(x_1)\succeq h(x_2)$)
output map $h:X\rightarrow \Outputset$, where $\Outputset$, the set of measurement or
output values, is a subset of some ordered space $\R^p$.

We also recall the following definition:
a system is {\em strongly monotone} if:
\[
\xi _1 \succ \xi _2 \quad \& \quad u_1 \succeq u_2 \quad\Rightarrow \quad x(t,\xi _1,u_1)
\gg x(t,\xi _2,u_2) \quad \forall \, t > 0. 
\]

It is often convenient to assume more about the steady-state convergence
properties of a monotone system. The following notion, first introduced in
\cite{monotone1} in slightly weaker form, will be useful in order to state our
main result. 

\bd{nondegenerate}
We say that a system admits a 
\emph{non-degenerate input to state (I/S) static characteristic}
$\kx (\cdot ): \mathcal{U} \rightarrow X$ if, for each constant input $u \in \mathcal{U}$,
there exists a unique globally asymptotically stable equilibrium $\kx(u)$ and
$\textrm{det} ( D_x f (\kx (u),u) ) \neq 0$.  
\ed

Notice that, for technical reasons, the property has been strengthened with
respect to the definition in \cite{monotone1} by assuming non-degeneracy of
the equilibria.
For systems with non-degenerate I/S characteristic, we also define their
\emph{input/output (I/O) characteristic} as the composition $h\circ \kx$.
 
Detecting if a system is monotone with respect to the partial order
induced by some positivity cone $K$, without actually having to
compute explicit trajectories of the system itself, is of course a very important task
in order to apply our results in any specific situation. 
Necessary and sufficient differential characterizations of
monotonicity are discussed in~\cite{monotone1}, where extensions to
systems with inputs and outputs are presented of some well-known criteria
previously only formulated for autonomous differential equations
(see~\cite{smith}).
For the sake of completeness we recall the differential characterization
proved in~\cite{monotone1}.  This characterization uses the concept of
contingent cones to subsets of Euclidean spaces.

\bt{inmonotone}
A finite-dimensional nonlinear systems of differential equations
$\dot {x} = f(x,u)$ with state-space $X$ and input-space $\mathcal{U}$ is
monotone, with respect to positivity cones $K$ on states and $\KU$ on inputs,
if and only if: 
\be{diffchar}
x_1 \succeq x_2 \textrm{ and } u_1 \succeq u_2 
\; \Rightarrow \; f(x_1,u_1) - f(x_2, u_2) \in \tc_{x_1 - x_2} K 
\ee
where $\tc_xK$ denotes the tangent cone to $K$ at the point $x$.
\et

An alternative characterization, also provided in~\cite{monotone1},
uses a generalization (to systems with inputs) of the concept of
quasi-monotonicity: the system~(\ref{cds}) is monotone if and only if
\[
\begin{array}{rcl}
\xi _1\succeq\xi _2,\,u_1\succeq u_2,\,\zeta \in K^*,
\mbox{ and }
\ip{\zeta }{\xi _1}=\ip{\zeta }{\xi _2}
\\
\Rightarrow \ip{\zeta }{f(\xi _1,u_1)}\;\geq \;\ip{\zeta }{f(\xi _2,u_2)}
\end{array}
\]
(it suffices to check this property for $\xi _1-\xi _2\in \partial K$),
where $K^*$ is the set of all $\zeta \in \R^n$ so that $\ip{\zeta }{k}\geq 0$ for all
$k\in  K$.

\subsection*{Orthant orders}

Any orthant $K$ in $\R^n$ has the form
\[
K^{(\varepsilon )} \;=\;
\{x\in \R^n \st (-1)^{\varepsilon _i}x_i\geq 0\,,\; i=1,\ldots ,n\}
\]
for some binary vector $\varepsilon =(\varepsilon _1,\ldots ,\varepsilon _n)\in \{0,1\}^n$.
Under appropriate changes of variables, one may often reduce the study of
monotone systems to the special case in which states, inputs, and outputs are
ordered with $K=$ the main orthant (all $\varepsilon _i=0$),
i.e. to the study of {\em cooperative systems}.
See~\cite{monotone1} for details.

\section{Statement of Main Results}

The property defined next was studied for linear systems in~\cite{farina}.

\bd{excitability} 
A system is \emph{excitable} if for any initial
condition $\xi $ and any pair of inputs $v,u$ with $v \succ u$
for almost all $t >0$,
the following holds:
\be{toprove}
x (t, \xi , v) \gg x(t,\xi ,u) \qquad \forall \, t > 0.
\ee  
It is {\em weakly excitable} if this is required for any 
pair of inputs $v,u$ with $v \gg u$.
\ed

The dual of excitability is also useful in the following discussion:

\bd{transparency}
A system is \emph{transparent} if for each 
  input $u$, and each pair of solutions 
$x(t, \xi _1, u)$, $x(t,\xi _2,u)$ with $\xi _1 \succ \xi _2$ we have
 $h(x(t,\xi _1,u)) \gg h(x(t,\xi _2,u))$ for all $t > 0$.
It is {\em weakly transparent} if the conclusion is that
$h(x(t,\xi _1,u)) \succ h(x(t,\xi _2,u))$ for all $t > 0$.
\ed

We prove in Section~\ref{sec-res1} this sufficient condition
for strong monotonicity of systems in unitary feedback:

\bt{firstresult}
Consider the unitary feedback interconnection of a system~(\ref{cds}), i.e.\
the system
\be{finsys}
\dot {x} =  f(x,h(x))
\ee
resulting when we
let $u=y$ 
and assume that inputs and outputs are ordered with respect to the same
positivity cone. 
The induced flow  is \emph{strongly} monotone provided that
(\ref{cds}) be monotone, excitable and transparent with either
excitability or transparency possibly holding in a weak sense.
\ets

Our main result will provide a global analysis tool for systems obtained
by positive feedback loops involving monotone systems. 
In~\cite{thomas}, R.\ Thomas conjectured that the existence of at least one
positive loop in the incidence graph is a necessary condition for the
existence of multiple steady states. 
Proofs of this conjecture were given in~\cite{gouze}, \cite{plahte},
\cite{snoussi}, and \cite{cinquin}, under different assumptions on the system
(the last reference provides the most general result, using a degree theory
argument).
However, the existence of positive loops is not sufficient, and our main
theorem deals precisely with this question.

The fixed points
of the I/O characteristic will play a central role in the statement of
the result.
In particular, we say that a map $k : \mathcal{U} \rightarrow  \mathcal{U}$ has
{\em non-degenerate fixed points} if 
for all $u\in \mathcal{U}$ with $k(u)=u$ we have that
$k^\prime(u)$ exists and $k^\prime (u)  \neq 1$. 

\bt{secondresult}
Consider a monotone, single-input, single-output ($m=p=1$, with
standard order) system, endowed with a
non-degenerate I/S and I/O static characteristic:
\be{feedback}
\begin{array}{rcl}
\dot {x} &=& f(x,u) \\
y &=& h(x).
\end{array}
\ee
Consider the unitary positive feedback interconnection $u=y$. Then the
 equilibria are in 1-1 correspondence with the fixed points of the
I/O characteristic.
Moreover, if $\ky$ has non-degenerate fixed points,  the closed-loop system is 
\emph{strongly monotone},
and all trajectories
are bounded, then for almost all initial conditions, solutions converge to
the set of equilibria of (\ref{feedback}) corresponding to inputs for which 
${\ky}^{\prime} ( u )<1$. 
\ets

This theorem is proved in Section~\ref{sec-res2}.
The fact that equilibria correspond to fixed points of the characteristic is
straightforward, but the global, and even local, stability statements are
nontrivial. 
The result is particularly useful when combined with the characterization of
strong monotonicity given in Theorem~\ref{firstresult}.

The next Section presents several graphical tests that are very useful in
checking the properties required by our Theorems.

\section{Graphical Conditions for Strong Monotonicity}

For the special case of positivity orthants, i.e.\ when the orders in each of
the input, state, and output spaces is defined by an orthant,
criteria may be formulated in terms of the \emph{incidence graph} of the
system.

Along similar lines to \cite{graphic1}, we associate to a given
system~(\ref{cds})
a signed digraph, with vertices 
$x_1, x_2 \ldots x_n$, $u_1, u_2, \ldots u_m$, $y_1, y_2 \ldots y_p$ and edges
constructed according to the following set of rules:

\noindent\textbf{Edges between $x$ vertices:}

\noindent
The graph is defined only for systems so that for any couple  
$1 \leq i, j \leq n$ of integers with $i \neq j$ one of the following
  rules apply:   
\begin{enumerate}
\item If $f^i (x,u)$ is strictly increasing with respect to $x_j$ for all
  $x, u \in X \times \mathcal{U}$  
then we draw a positive edge $e_{ij}^x$ directed from vertex $x_j$ to $x_i$.
 \item  If $f^i (x,u)$ is strictly decreasing as a function of $x_j$ for all
   $x, u \in X \times \mathcal{U}$ 
then we draw a negative edge $e_{ij}^x$ directed from vertex $x_j$ to $x_i$.
\item Otherwise, $\frac{\partial f^i}{\partial x_j} = 0$ for all $x,
  u$ and no edge from $x_j$ to $x_i$ is drawn.
\end{enumerate}

\noindent\textbf{Edges between $u$ and $x$ vertices:}

\noindent
The graph is defined only for systems so that for any couple of 
integers $i$, $j$ with  $1 \leq i \leq n$ and $1 \leq j \leq m$ one of the following rules
apply:
\begin{enumerate}
\item If $f^i (x,u)$ is strictly increasing as a function of $u_j$ for all
  $x,u \in X \times \mathcal{U}$ 
then we draw a positive edge ${e}_{ij}^u$ directed from vertex $u_j$ to $x_i$.
\item If $f^i (x,u)$ is strictly decreasing as a function of $u_j$ for  all $x, u
  \in X \times \mathcal{U}$
then we draw a negative edge ${e}_{ij}^u$ directed from vertex $u_j$ to $x_i$.
\item Otherwise $\frac{\partial f^i}{\partial u_j} = 0$ for all $x, u$ and no edge from $u_j$ to $x_i$ is drawn.
\end{enumerate}

\noindent\textbf{Edges between $x$ and $y$ vertices:}

\noindent
The graph is defined only for systems so that for any couple of
integers $i$, $j$ with  $1 \leq i \leq p$ and $1 \leq j \leq n$ one of the following rules apply:   
\begin{enumerate}
\item If $h^i (x)$ is strictly increasing as a function of $x_j$ for  all
  $x \in X$ 
then we draw a positive edge ${e}_{ij}^y$ directed from vertex $x_j$ to $y_i$.
\item If $h^i (x)$ is strictly decreasing  as a function of $x_j$ for  all $x \in X $
then we draw a negative edge $e_{ij}^y$ directed from vertex $x_j$ to $y_i$.
\item Otherwise, $\frac{\partial h^i}{\partial x_j} = 0$ for all $x
  \in X$ and no edge from $x_j$ to $y_i$ is drawn.
\end{enumerate}

When there is no risk of confusion, we just write from now on just
``$e_{ij}$'' to refer to an edge of the type $e_{ij}^x$, $e_{ij}^u$, or
$e_{ij}^y$.

Under this convention, a directed path $\mathcal{P}$ is a finite sequence of vertices, $v_{n_0}, v_{n_1} \ldots v_{n_L}$, such that
each vertex appears at most once in the sequence and $e_{ij}$ is an edge whenever $v_j, v_i$ appear
consecutively in the path. The integer $L$, is called the length of the path and it is denoted by  $L(\mathcal{P})$.
By $\mathcal{P}_i$, we denote the $v_{n_i}$, the $i+1$-th, vertex of the path $\mathcal{P}$. 
A \emph{cycle}, not necessarily directed, is a finite sequence of vertices $v_{n_0}, v_{n_1} \ldots v_{n_L}$ such that
$v_{n_0} = v_{n_L}$ and the constraint that either $e_{ij} $ or $e_{ji}$ is an edge whenever $v_i$ and $v_j$ appear consecutively in the
cycle.
The sign of a cycle is defined as the product of the signs of the edges
comprising it, and the sign of a path is defined to be the product of the signs of
its edges.

One of the main results in \cite{graphic1} is that an autonomous system (no
inputs) is monotone with respect to some orthant if and only if its associated
graph does not contain any negative cycles.  An analogous
result (basically with the same proof, which therefore we omit),
holds for controlled systems:

\bp{monotonecontrol}   
A system (\ref{cds}) which admits an incidence graph according to the
above set of rules is monotone with respect to some orthants $K$,
$\KU$ and $\KY$ if and only if its graph
does not contain any negative cycles.
\ep

\br{whythisclass}
We remark that in this set-up we deliberately restricted the class of
systems for which the incidence graph is defined. In \cite{graphic1}
in fact the milder requirement that  
$\frac{ \partial f^i}{\partial  x_j} \geq 0$ for all $x$ 
together with $\frac{ \partial f^i}{\partial x_j} >0$
for some $x$ is asked for in order to draw an edge between vertices
$x_i$, $x_j$. 
This more general notion of incidence graph is however much more
cumbersome to deal with if we want to give conditions for strong
monotonicity of a system. 
\er

This definition of incidence graph also provides the right set-up for 
easy geometrical characterizations of excitability and transparency;
see~\cite{farina2} for systems with no inputs and outputs:

\bt{excitability2}
A monotone system which admits an incidence graph is excitable provided that
each $x_i$ is reachable through a directed path from any $u_j$,
and it is weakly excitable provided that
each $x_i$ is reachable through a directed path from some $u_j$,
\ets

It is worth pointing out that for the special case of positive {\em linear}
systems the above results are proven in \cite{farina}.

Theorem \ref{excitability2} is proved in an Appendix.

Similarly, we have:

\bt{transparency2}
A monotone system which admits an incidence graph is (weakly) transparent provided that
directed paths exist from any $x_j$ to any
(at least one) output vertex $y_i$ .
\ets

The proof of Theorem \ref{transparency2} is analogous to that of
Theorem~\ref{excitability2}, and is sketched in an Appendix.

\section{Proof of Theorem~\protect{\ref{firstresult}}}
\label{sec-res1}

By Theorem \ref{inmonotone}, we know that
\be{monchar}
 x_1 \succeq x_2 \; \& \;  u_1 \succeq u_2 \; \Rightarrow 
f(x_1,u_1) - f(x_2,u_2) \in \tc_{x_1 - x_2} K\,
\ee
where $K$ is the positivity cone relative to the order $\succeq$.
Let us first show monotonicity of the feedback loop system. Recall
that $h$ is a monotone map, i.e.:
\be{implication1}
x_1 \succeq x_2 \; \Rightarrow \; h(x_1) \succeq h(x_2) 
\ee  
Therefore, if we combine (\ref{monchar}) with (\ref{implication1}) and
we let $u_1 = h (x_1)$ and $u_2 = h(x_2)$ we obtain
\be{implication2}
x_1 \succeq x_2 \; \Rightarrow f(x_1,h(x_1)) - f(x_2,h(x_2)) \in
\tc_{x_1 - x_2} K  
\ee
which, by Theorem 1 in \cite{monotone1} is equivalent to
monotonicity of the closed-loop system:
\be{clsys}
 \dot {z} = f(z,h(z)).
\ee
In particular then, if we denote by $z(t,\xi )$ the solutions  of
(\ref{clsys}) we have as a consequence of monotonicity:
\be{implication3}
 \xi _1 \succeq \xi _2 \; \Rightarrow h ( z (t,\xi _1) ) \succeq h ( z (t,
\xi _2 )) \qquad \forall \; t \geq 0. 
\ee
Exploiting the fact that $z ( t , \xi ) = x ( t, \xi , h ( z (\cdot ,\xi )))$ and (weak)
strong transparency of (\ref{cds}) we obtain: 
\begin{eqnarray}
\label{strongm}
\xi _1 \succ \xi _2 \;& \Rightarrow & h ( z ( t , \xi _1))= h(x(t,\xi _1, h
(z(\cdot ,\xi _1))) ) \gg (\succ) \;  h ( x(t,\xi _2, h(z(\cdot ,\xi _1)))) \nonumber \\
& \succeq & h( x(t,\xi _2, h(z(\cdot ,\xi _2)))) = h(z(t,\xi _2)) \qquad
\forall \, t >0.
\end{eqnarray}
Finally, by (\ref{strongm}) and weak (strong) excitability:
\begin{eqnarray}
\xi _1 \succ \xi _2 & \Rightarrow &
h(z(t,\xi _1)) \gg (\succ) \;  h(z(t,\xi _2)) \nonumber \\
& \Rightarrow &  z(t,\xi _1) = x(t,\xi _1,h(z(\cdot ,\xi _1))) \gg x(t,\xi _2,h(z(\cdot ,\xi _2)))
= z(t,\xi _2) \qquad \forall \; t >0
\end{eqnarray}
as desired.
\qed

\section{Monotone Linear Systems}

We recall next some basic facts about linear monotone systems which
will be of interest in the discussion of the main result.

\bt{linearmonotone}
Let us consider the following finite dimensional linear system:
\be{linearsys}
\dot {x} = A x + B u, \qquad y = C x.
\ee
with $x \in \R^n$, $u \in \R^m$, $y \in \R^p$
and assume the state, input and output space equipped with some partial 
orders induced by the positivity cones $\KX$, $\KU$ and $\KY$ respectively. 

System (\ref{linearsys})  is a monotone control system with respect to the partial
orders specified above if and only if:
\begin{enumerate}
\item $\KX$ is positively invariant for the autonomous system $\dot {x} = Ax$;
\item $B \KU \subseteq \KX$;
\item $C \KX \subseteq \KY$.
\end{enumerate} 
\ets
\bpr 
By the characterization of monotonicity in Theorem \ref{inmonotone}, a system
is monotone if and only if: 
\be{char}
x_1 \succeq x_2 \; \& \; u_1 \succeq u_2 \quad \Rightarrow \quad A (x_1-x_2) + B (u_1 - u_2) \in \tc_{x_1-x_2} \KX,
\ee
and the output map is monotone, i.e.:
\be{char2}
x_1 \succeq x_2 \Rightarrow C x_1 \succeq C x_2.
\ee 
In terms of positivity cones  and denoting $\tilde{x}:= x_1 -x_2$ and $\tilde{u}=u_1 - u_2$, 
conditions (\ref{char}) and (\ref{char2}) are equivalent to:
\be{char3}
\tilde{x} \in \KX \; \& \; \tilde{u} \in \KU \; \Rightarrow A \tilde{x} + B \tilde{u} \in \tc_{\tilde{x}} \KX
\ee
and:
\be{char4}
\tilde{x} \in \KX \Rightarrow C \tilde{x} \in \KY.
\ee
Condition (\ref{char4}) is clearly equivalent to assumption 3).
Condition (\ref{char3}) can be further decomposed by first taking arbitrary $\tilde{x}$ and fixing 
$\tilde{u}=0$ and then $\tilde{x}=0$ and arbitrary $\tilde{u}$. Condition (\ref{char3}) therefore implies (and is in fact
equivalent to, as we shall see later):
\be{char5}
\tilde{x} \in \KX \; \Rightarrow \; A \tilde{x} \in \tc_{\tilde{x}} \KX
\ee
and:
\be{char6}
\tilde{u} \in \KU \; \Rightarrow \; B \tilde{u} \in \tc_0 \KX = \KX.
\ee
The converse implication just follows by recalling that tangent cones of a convex set are closed under sums (since they are convex cones) and
the following inclusion holds: $\KX \subseteq \tc_{\tilde{x}} \KX$ for any $\tilde{x} \in \KX$. 
Condition (\ref{char6}) is clearly assumption 2). Whereas condition (\ref{char5}) is the well-known characterization of
positive invariance of $\KX$ under the flow $\dot {x}=Ax$.
\epr
\bc{easyconsequence}
The impulse response  of a finite-dimensional, monotone, linear system (with respect to
positive impulses)  is a positive signal in output space:
\[ C e^{A t } B \KU \; \subseteq  \; \KY. \]
\ec

The following fact, reviewed in an appendix, is a straightforward consequence
of the Perron-Frobenius (Krein-Rutman) Theorem (see \cite{berman} pp.\ 6-8):
 
\bl{Perronfrob}
Assume that the linear system $\dot {x}=A x$ admits a positively invariant
convex (and proper) cone $K$.
Then, there exists a dominant real eigenvalue $\lambda $ (i.e.\ an
eigenvalue so that $\textrm{Re} [ \lambda _i ] \leq \lambda $ for all $i
\in 1, 2, \ldots n$), and a corresponding
nonnegative
eigenvector $v_\lambda $ 
(positive and unique up to a positive multiple if $A$ is irreducible) 
satisfies
$v_\lambda \in K$. 
\els

\br{gains} It is worth pointing out that
for asymptotically stable single-input single-output monotone systems, the
condition $h(t) \geq 0$, implies that the 
$L_\infty \rightarrow L_\infty $ induced gain equals the steady state gain.
Recall that the steady-state gain of a linear system is just the slope of
its $I/O$ static characteristic. 
The $L_\infty \rightarrow L_\infty$ induced gain is instead defined as:
\[ \gamma _\infty : = \sup_{ u \neq 0 } \frac{ \| y \|_\infty }{\| u \|_\infty}
\]
where $y(t)=y(t, 0, u)$. It is well known (see \cite{doyle}, pg.\ 16) that
$\gamma _\infty$ equals the $L_1$ norm of the impulse response. Thus, 
\[ \gamma _{\infty}  = \int_0^{+\infty} | h(t) | \, dt
= \int_0^{+\infty} h(t) \, dt = - C A^{-1} B 
. \] 
This last quantity equals ${\ky}^\prime (u)$ for any $u$, for linear systems.
When the linear system in question is obtained by linearizing a nonlinear
system about a steady state corresponding to an input $u_0$, it equals 
${\ky}^\prime (u_0)$, where $\ky$ is the I/O characteristic of the original
nonlinear system.
\er

The next technical lemma will be useful in order to study nonlinear
monotone systems by linearizing the flow around an equilibrium
position:

\bl{linearization}
Let $f: X \times \mathcal{U} \rightarrow \R^n$ be a
$\mathcal{C}^1$ vector-field. Let $f(\bar{x},\bar{u})=0$ for some $\bar{x}
\in X$ and $\bar{u} \in \mathcal{U}$. If the flow induced by $f$ is
monotone with respect to some positivity cone $K$, the same holds true
for the linearization at $(\bar{x}, \bar{u})$:
\be{linsys}
\begin{array}{rcl}
\dot {z} &=& \left . \frac{ \partial f}{\partial x} \right |_{x=\bar{x},
u= \bar{u} } z + \left . \frac{ \partial f}{\partial u} \right
|_{x=\bar{x}, u=\bar{u} } v  \\
w & = & \left . \frac{\partial h}{\partial x} \right |_{x = \bar{x} } z .
\end{array} 
\ee  
\els
\bpr By one the  results in \cite{monotone1}, a system is monotone
with respect to the positivity cones $K$ (for states) and $\KU$ (for inputs) if and only if:
\be{charact}
x_1 \succeq x_2, \; u_1 \succeq u_2 \Rightarrow f(x_1,u_1) - f(x_2,u_2
) \in \tc_{x_1 - x_2} K.
\ee
Let $z \in K$, $v \in \KU$ be arbitrary and, for any $\varepsilon >0$,
$x_\varepsilon := \varepsilon z + \bar{x}$,
$u_{\varepsilon } = \varepsilon v + \bar{u}$.
By (\ref{charact}) applied with $x_1 = x_{\varepsilon }$, $x_2 = \bar{x}$,
$u_1=u_\varepsilon $, and $u_2=\bar u$,
\be{tendingtozero}
f(x_{\varepsilon },u_{\varepsilon }) / \varepsilon \in \tc_{\varepsilon z}
(K) = \tc_{z} K.
\ee
By letting $\varepsilon $ tend to $0$ and exploiting closedness of the
tangent cone we have:
\be{forthelinear}
z \succeq 0, \; v \succeq 0 \; \Rightarrow 
  \left . \frac{ \partial f}{\partial x} \right |_{x=\bar{x},
u= \bar{u} } z + \left . \frac{ \partial f}{\partial u} \right
|_{x=\bar{x}, u=\bar{u} } v \;  \in \; \tc_{z} K.
\ee
Let, for simplicity $A = \left .  \frac{ \partial f}{\partial x} \right |_{x=\bar{x},
u= \bar{u} }$ and $B = \left . \frac{ \partial f}{\partial u} \right
|_{x=\bar{x}, u=\bar{u} }$.
By linearity, there follows easily:
\be{incremental}
z_1 \succeq z_2 \; , \; v_1 \succeq v_2 \; \Rightarrow 
 (A z_1 + B v_1) - (A z_2 + B v_2) \in \tc_{z_1 - z_2} K.
\ee
This concludes the proof of the claim, by exploiting once more the
characterization of monotonicity in \cite{monotone1}.
\epr

We remark that for the special case of $K$, $\KU$ being positive
orthants the result was already proved in Section 8, \cite{monotone1}.

\bl{nondegenerateimpliehyperbolic}
Consider a monotone system with a non-degenerate I/S static characteristic $\kx (\cdot )$. For each
$u \in \mathcal{U}$ the corresponding equilibrium $\kx(u)$ is hyperbolic.
\els
\bpr 
By Lemma \ref{linearization} the linearized system at the equilibrium is
monotone. 
Therefore it admits a real dominant eigenvalue $\lambda $. By asymptotic stability
of the nonlinear system and non-degeneracy, $\lambda <0$. 
Thus for all eigenvalues $\lambda _i$ of $D_x f (\kx (u),u)$ we have 
$\textrm{Re}[\lambda _i] \leq \lambda  < 0$ which completes the proof of our claim.
\epr

The following key fact establishes a relation between steady-state responses
and stability under unity-feedback, for monotone linear systems.

\bl{key-linear}
Suppose that the linear system $\dot z=Az+Bu$, $y=Cx$ is monotone, where inputs
and outputs are scalar and are endowed with the standard order in $\R$, the
matrix $A$ is Hurwitz (all eigenvalues have negative real parts), and
$CA^{-1}B \not=  -1$. 
Then, the following two properties hold:
\ben
\item
$CA^{-1}B< -1$ if and only if every eigenvalue of $A+BC$ has negative real
part.
\item
$CA^{-1}B> -1$ if and only if there is an eigenvalue of $A+BC$ with positive
real part.
\een
\els

\bpr
We start by noticing that the closed-loop system $\dot z=(A+BC)z$ is monotone,
since monotonicity is preserved under positive feedback, as shown in the first
part of the proof of Theorem~\ref{firstresult}.
Therefore, the matrix $A+BC$ admits a real dominant eigenvalue $\bar{\lambda }$,
i.e.\ an eigenvalue so that $\bar{\lambda }\geq \textrm{Re} [ \lambda _i ]$
for all eigenvalues $\lambda _i$ of $A+BC$,
and there is a corresponding
eigenvector: 
\be{eds-eq1}
( A+BC ) \bar{v} = \bar{\lambda } \bar{v}
\ee
with $\bar{v} \in \KX=K$.
By choice of $\bar{\lambda }$, the condition $\bar{\lambda }>0$ (respectively $\bar{\lambda }<0$)
is equivalent to there existing some eigenvalue of $A+BC$ with positive real
part (respectively, all eigenvalues have negative real part).

We now multiply both sides of~(\ref{eds-eq1}) by $C A^{-1}$, and obtain:
\be{multiplied}
\bar{\lambda } \, ( C A^{-1} \bar{v}) \, = \, ( C \bar{v} ) \, [ 1 + C A^{-1}
  B ] \,.
\ee
Note that $\bar{\lambda }\not= 0$.  
Otherwise, if $\bar{\lambda }=0$, Equation~(\ref{multiplied}) together with
the fact that 
$CA^{-1}B \not=  -1$,
would imply that $C\bar{v}=0$, and hence
$A\bar{v} = ( A+BC ) \bar{v} = \bar{\lambda } \bar{v}=0$, which would mean that $A$ is
singular, contradicting the nondegeneracy assumption on steady states.
Thus, we know that $\bar{\lambda }\not= 0$ and that 
$1+CA^{-1}B \not=  0$.
So we must show that $1+CA^{-1}B>0$ iff $\bar{\lambda }<0$.

We know that $C\bar{v}\geq 0$, by Property 3 in Theorem~\ref{linearmonotone}, and
\be{inequality}
C A^{-1} \bar{v} = - \int_{0}^{+\infty} C \underbrace{e^{At}
\bar{v}}_{\in K} \, dt \leq  0,
\ee
where the integral in (\ref{inequality}) converges as $A$ is Hurwitz.
If $C\bar{v}>0$, then $CA^{-1}\bar{v}<0$,
and hence~(\ref{multiplied}) gives that $1+CA^{-1}B>0$ iff $\bar{\lambda }<0$, as
wanted.  So, we must only treat the case $C\bar{v}=0$.  We do this next, by
means of a perturbation argument.

Since $K$ is a pointed cone, there is some vector $p\in \R^n$ with the property
that $\ip{p}{v}>0$ for all $v\in K\setminus\{0\}$ (see e.g.~\cite{lewis-borwein},
Theorem 3.3.15).
For each $\varepsilon >0$, let $C_\varepsilon :=C+ \varepsilon p'$.
Note that $C_\varepsilon  v > 0$ for all $v\in K\setminus\{0\}$, because of the choice of
$p$ and because $Cv\geq 0$.
Moreover, by continuity on $\varepsilon $, for all $\varepsilon $ small enough (assumed from now
on), $C_\varepsilon A^{-1}B \not=  -1$.
Thus, we may apply the previous proof to the system described by $(A,B,C_\varepsilon )$
(note that the vector $\bar{v}$ picked in the proof belongs to
$K\setminus\{0\}$).
We conclude that, for all small $\varepsilon >0$, 
\[
1+C_\varepsilon A^{-1}B>0 \;\Longleftrightarrow\; \bar{\lambda }_\varepsilon <0
\]
where $\bar{\lambda }_\varepsilon $ is the dominating eigenvalue of $A+BC_\varepsilon $.
We have that $1+C_\varepsilon A^{-1}B\rightarrow 1+C A^{-1}B$ and, by continuity of eigenvalues on
matrix entries (e.g., Appendix A.4 in~\cite{mct}), also
$\bar{\lambda }_\varepsilon \rightarrow \bar{\lambda }$ as $\varepsilon \searrow0$.
The result then follows by taking limits and recalling that we know that
$\bar{\lambda }\not= 0$ and $1+CA^{-1}B \not=  0$.
\epr

\section{Proof of Theorem~\protect{\ref{secondresult}}}
\label{sec-res2}

Let $\kx: \mathcal{U} \rightarrow X$ denote the I/S static
characteristic and let $ \bar{u} $ be any solution of $u = h ( \kx (u) )$.
Clearly, $f(\kx (\bar{u}), h(\kx (\bar{u}) ) )  = f(\kx (\bar{u} ), \bar{u} ) = 0$
and therefore $\bar{x} := \kx ( \bar{u} )$ is an equilibrium
of the closed-loop system. Conversely, let $\bar{x} $ be an
equilibrium; the corresponding output value satisfies $\bar{y} = h
(\bar{x})$. As in closed-loop $u=y$, we have $\bar{x}= \kx ( \bar{y}
)$. Thus $\bar{y} = h ( \kx ( \bar{y} ))$, as desired.

The characteristic $\kx$ is a differentiable function.
Indeed, we have that $\kx(u)$ solves $f( \kx(u),u) =0$, and
the nondegeneracy assumption says that the partial derivative of $f(x,u)$ with
respect to $x$ is invertible at $\kx(u)$, for each $u$; by the Implicit
Mapping Theorem, it follows that $\kx$ is differentiable.
Moreover, we can compute its derivative by differentiating: 
\[
\frac{ \partial f}{\partial x} ( \kx (u) , u) \; 
{\kx}^{\prime} (u) + \frac{ \partial f}{\partial u} (\kx (u),u) \;=\; 0\,.
\]
Evaluating the above expression at $u = \bar{u}$ yields 
$k^{\prime}_x (\bar{u} ) = - A^{-1} B$
and so 
\[
k^{\prime}_y(\bar{u})  
\;=\; \frac{ \partial h }{\partial x} ( \kx (u) ) \; k^{\prime}_x (u)\;=\;
-CA^{-1} B\,,
\]
where
$A$, $B$, and $C$ are defined as:
\[A = \left .  \frac{ \partial f}{\partial x} \right |_{x=\kx(\bar{u}), u= \bar{u} },\quad
B = \left . \frac{ \partial f}{\partial u} \right |_{x=\kx (\bar{u} ), u=\bar{u} },\quad
C = \left . \frac{ \partial h}{\partial x} \right |_{x=\kx (\bar{u} ) }
\]
and $A^{-1}$ exists by non-degeneracy of the I/S characteristic.
(Note that this gives,, in particular, that the $\mathcal{L}^\infty$ induced
gain of the linearized system (\ref{linsys}) is
$\gamma _{\infty} = k^{\prime}_y ( \bar{u} )$, by
Remark \ref{gains}.)

Next, we turn to the relation between stability and the slopes
of the $I/O$ characteristic at equilibria.
The closed-loop linearized system which arises by linearizing the
nonlinear system (\ref{feedback}) together with the unitary feedback
interconnection $u=y$
is precisely the same as the system that results if we first linearize
(\ref{feedback}), obtaining
$\dot z=Az+Bu$, $y=Cz$
(which is itself monotone by virtue of Lemma \ref{linearization}), and then
apply unitary feedback to obtain $\dot {z} = (A+BC) z$.
Note that $A$ is a Hurwitz matrix, by Lemma~\ref{nondegenerateimpliehyperbolic}.
Also, $CA^{-1}B\not= -1$, because $k^{\prime}_y ( \bar{u} )\not= 1$
(nondegenerate characteristic).
Thus, we may apply Lemma~\ref{key-linear}.

In particular, equilibria with $k^{\prime}_y (\bar{u}) < 1$ are
locally asymptotically stable and equilibria with $k^{\prime}_y (
\bar{u} ) > 1$ have a nontrivial  unstable manifold.
By Hirsch's Theorem on generic convergence of strongly monotone
flows (see \cite{Hirsch}, Section 7), for almost all initial conditions,
solutions will converge to the set of equilibria. 
Moreover, by Remark~\ref{basinofattraction} below, the stable manifolds of
(exponentially) unstable equilibria have zero-measure. Therefore, for almost
all initial conditions, solutions converge to the set of points where
$k^{'}_y(\bar{u}) < 1$.
This completes the proof of our result.  
\qed

\br{uniqueequilibrium} It is worth pointing out that, whenever the
equilibrium in Theorem \ref{secondresult} is unique, convergence to
the equilibrium is global 
under mild assumptions of convexity and location of omega limit sets in
the interior of the state-space; see
Theorem 3.1 of \cite{smith}. 
\er 
\br{freqdomain} An alternative proof, based on frequency domain considerations,
of the connection between stability of the closed-loop equilibrium and
the I/O characteristic is provided next.

Consider the transfer function
\[
w(s) = \int_0^\infty h(t) e^{-st} dt
\]
of a strictly proper linear system, and let
\[
w_{\rm cl}(s) = \frac{w(s)}{1-w(s)}
\]
be the transfer function of the associated unity-feedback closed-loop system.
Suppose:
\begin{enumerate}
\item
$h$ is integrable (so, $w$ has no real nonnegative poles);
\item
$h(t)\geq 0$ for all $t\geq0$ (and is not identically zero);
\item
$w(0)\not= 1$ (transversality condition).
\end{enumerate}
Then:
\begin{itemize}
\item[(a)]
  there exists a positive real pole of $w_{\rm cl}$
if and only if
  $w(0)>1$;
\item[(b)]
  every real pole of $w_{\rm cl}$ is negative
if and only if
  $w(0)<1$.
\end{itemize}

\emph{Proof:}
By the first assumption, $w(\lambda )$ is a continuous 
(real-valued) function for $\lambda \geq 0$.

Furthermore, $h(t)\geq 0$ for all $t\geq0$ and not identically zero
implies that 
$w'(\lambda ) = -\int_0^\infty  h(t) t e^{-\lambda t} dt <0$ for all $\lambda $, 
so $w$ is a strictly decreasing function of $\lambda $.

Nonnegative real poles of $w_{\rm cl}$ are exactly those 
$\lambda \geq 0$ such that $w(\lambda )=1$.

If $w(\lambda )=1$ for some $\lambda >0$ then 
the strict decrease of $w$ implies that
$w(0)>1$.
Conversely, suppose that $w(0)>1$.
By strict properness, $w(\lambda )\rightarrow 0$ as $\lambda \rightarrow +\infty $.
Thus there is some $\lambda >0$ such that $w(\lambda )=1$.
This proves (a).

The first conclusion may be restated as:
``every pole of $w_{\rm cl}$ is $\leq 0$ if and only if $w(0)\leq 1$''
so, since we know in addition that $w(0)\not= 1$, this is the same as
requiring that every real pole is (strictly) negative.
Thus (b) holds too.
\er

\br{basinofattraction}
Stable manifolds of (exponentially) unstable equilibria have zero-measure.
In the non-necessarily hyperbolic case, this fact is an easy consequence of
Theorem 2.1 in \cite{deLave} (modified as discussed in the remarks following Theorem
2.1, including the choice of suitable norms and the multiplication by a
``bump'' function, after a linear change of coordinates, and specialized to
$r=1$, and applied to time-1 maps).
\er

\br{basins} A precise characterization of the basin of attraction of
each asymptotically stable equilibrium is of course not possible in
general; on the other hand, it is a straightforward consequence of 
monotonicity of the $I/S$ characteristic that equilibria are ordered,
$e_1 \prec e_2 \prec e_3$. It therefore makes sense to speak about
intervals $[e_1, e_2] := \{ x \in X : e_1 \preceq x \preceq e_2 \}$.
Again, it is a straightforward consequence of monotonicity that
intervals $[e_1,e_2]$ with $e_1$, $e_2$ equilibria 
are positively invariant. This allows to give estimates of the basin
of attraction of each equilibrium. In the case of $3$ equilibria for
instance, with $e_1 \prec e_2 \prec e_3$ and $e_1, e_3$ asymptotically stable,
$e_2$ unstable, we can conclude that
$\{ x : x \ll e_2 \} \subset \mathcal{A}_1$
and $\{ x: x \gg e_2 \} \subset \mathcal{A}_3$.
Similar considerations, based on empirical evidence,
are made for instance in \cite{bhalla}. It is therein pointed-out how
the unstable equilibrium plays the role of a threshold.   
\er

\section{Examples}

\begin{figure}[htl]
\begin{center}
\includegraphics[width=6cm]{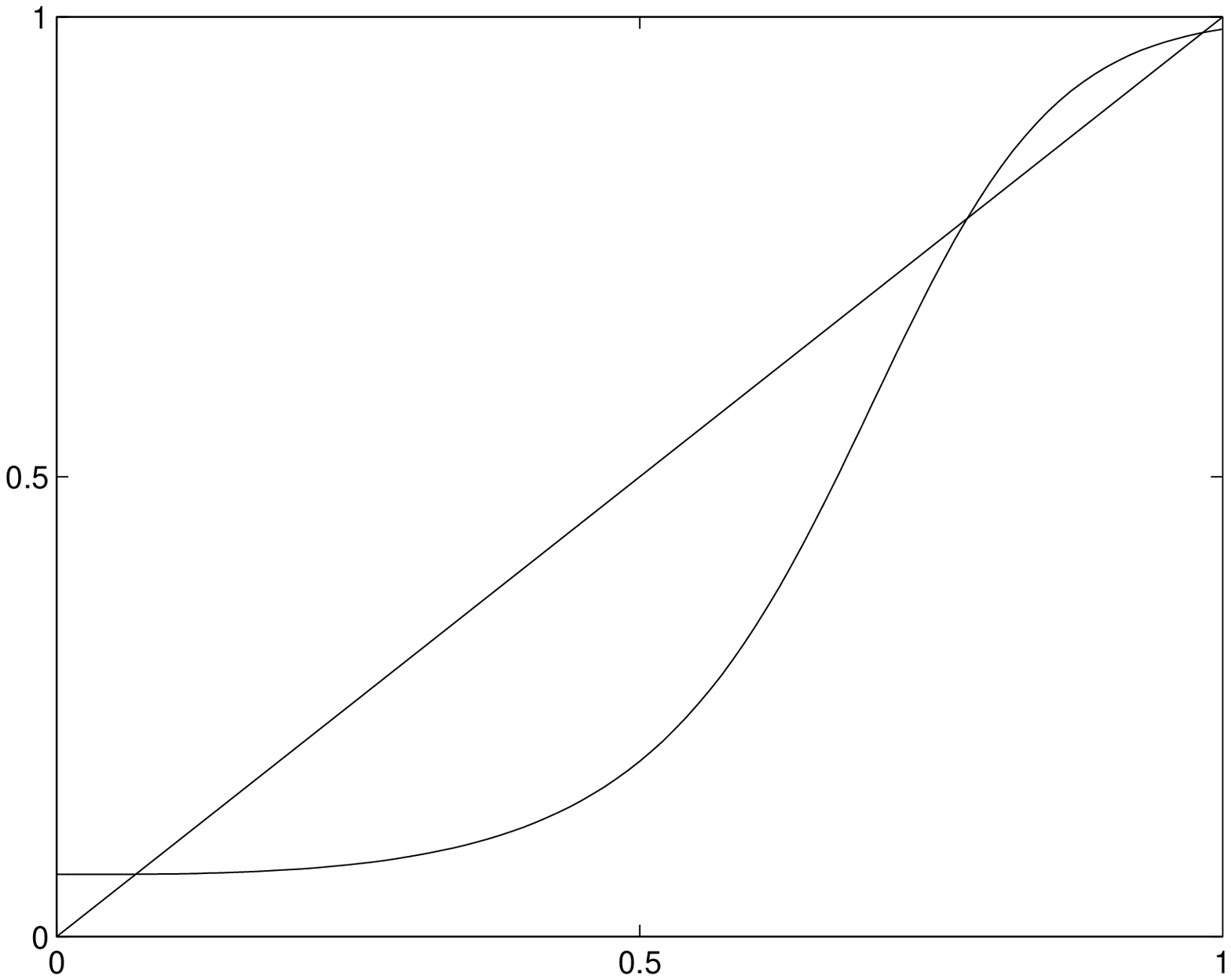}
\ \ \ 
\includegraphics[width=6cm]{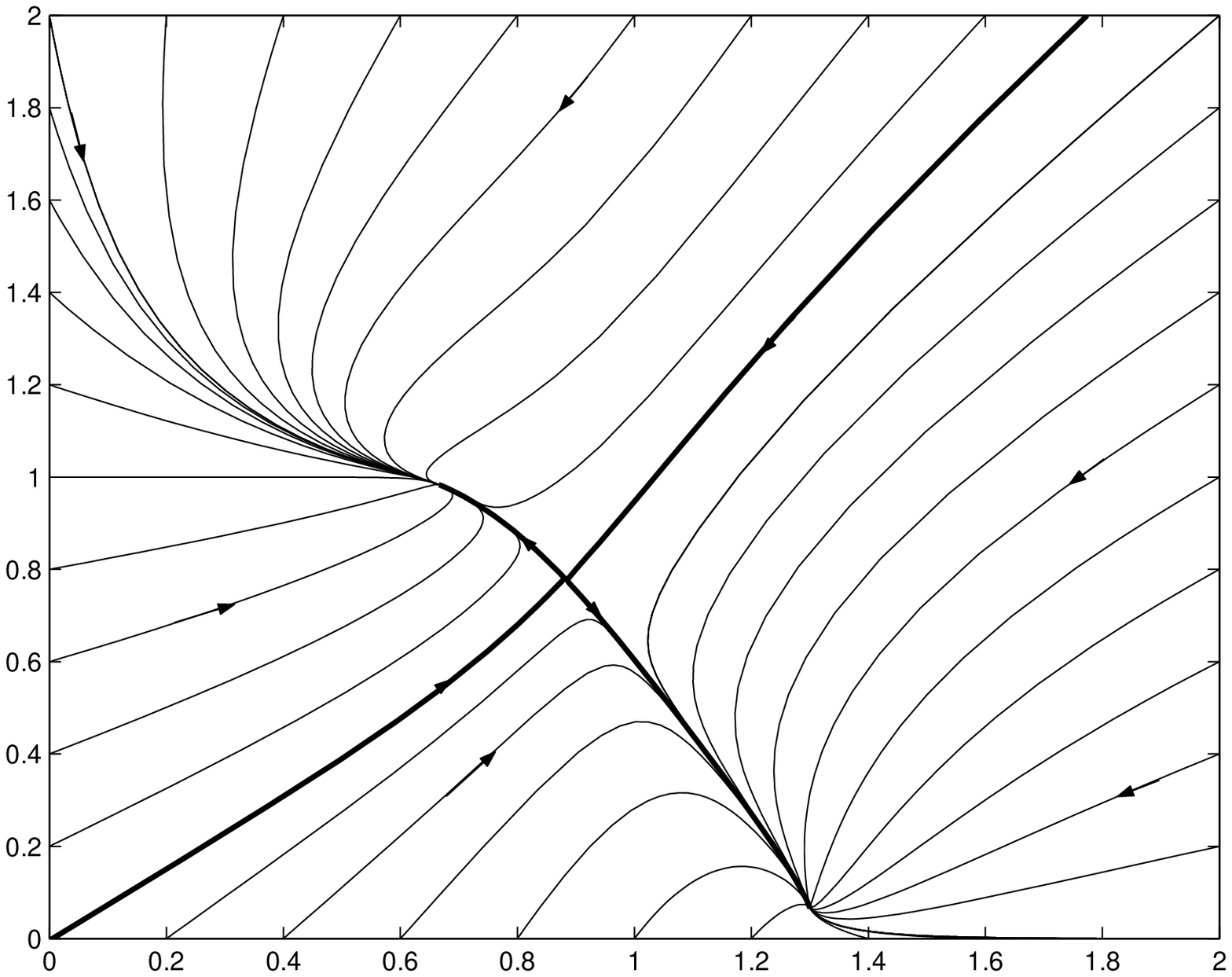}
\end{center}
\caption{I/O characteristic and phase plane.
Horizontal axis is $u$ (resp., $x_1$) and vertical axis is $x_2$.}
\label{iochar}
\end{figure}

A typical situation for the application of Theorem
\ref{secondresult} is when a monotone system with a well-defined I/O 
characteristic of sigmoidal shape is closed under unitary feedback.
If the sigmoidal function is sufficiently steep, this configuration is
known to yield $3$ equilibria, $2$ stable and $1$ unstable.  
In biological examples this might arise when a feedback loop
comprising any number of positive interactions and an even number of
inhibitions is present (no inhibition at all is also a situation which might
lead to the same type of behavior). This is a well-known principle in
biology. 
One of its simplest manifestations is the so called ``competitive
exclusion'' principle, in  which one of two competing species completely
eliminates the other, or more generally, for appropriate parameters the
bistable case in which they coexist
but the only possible equilibria are those where either one of the species
is strongly inhibited.
As a simple example, consider the system described
in \cite{nature}, used there to describe a model of gene expression.
The systems equations are as follows:
\be{ecoli}
\begin{array}{rcl}
\dot {x}_1 &=& \frac{\alpha _1}{1+ x_2^\beta } - x_1 \\ 
\dot {x}_2 &=& \frac{\alpha _2}{1+ x_1^\gamma } - x_2
\end{array}
\ee
where $\alpha _1,\alpha _2,\beta ,\gamma $ are some positive constants.
This can be seen as the unitary feedback closure of:
\be{ecoli2}
\begin{array}{rcl}
\dot {x}_1 &=& \frac{\alpha _1}{1+ u^\beta } - x_1 \\ 
\dot {x}_2 &=& \frac{\alpha _2}{1+ x_1^\gamma } - x_2 \\
y & = & x_2. 
\end{array}
\ee
Equation (\ref{ecoli2}) is a monotone dynamical system with respect
to the order induced by the positivity cone $K := \R_{\leq 0} \times \R_{\geq
0}$. 
It is straightforward by a cascade argument to see that the system is
endowed with the following static I/S characteristic:
\[ \kx (u) = \left [ \begin{array}{c} \frac{\alpha _1}{1+ u^\beta } \\
\frac{\alpha _2 (1 + u^\beta )^\gamma }{(1+u^\beta )^\gamma + \alpha _1^\gamma }
\end{array} \right ] .
\]
In Fig.~\ref{iochar} we plotted the I/O static characteristic for 
$\alpha _1=1.3$, $\alpha _2=1$, $\beta =3$ and  $\gamma =10$. 
(The value $\gamma =10$ was chosen only in order to help visualize the sigmoidal
form of the characteristic, and similar results hold for a smaller and more
biologically realistic constant.)
As confirmed by a sketch of the phase plane, for almost all initial conditions,
trajectories converge to the equilibria where the derivative condition is
satisfied.

Of course, the interest of our results is in the high-dimensional case in
which phase-plane techniques cannot provide the result, and we turn to such an
example next.  However, let us note that, for the special case of
two-dimensional systems, our techniques are very close to those
of~\cite{Adler}.  In fact, even the 4-dimensional example of a two-repressor
system with RNA dynamics, treated in~\cite{Adler} (Appendix I) in an ad-hoc
manner, can be shown to be globally bistable as an immediate application of our
techniques.

We now turn to a less trivial example where our tools may be applied.
(A different example, involving cascades of systems of this type, and with
comparisons with experimental data, is treated in~\cite{angeli-ferrell}.)
Consider the following chemical reaction, involving various forms of a protein
E:
\[
\mbox{E}_1
\;\;\arrowschem{\mbox{\small U}}{}\;\;
\mbox{E}_2
\;\;\arrowschem{\mbox{\small U}}{}\;\;
\ldots 
\;\;\arrowschem{\mbox{\small U}}{}\;\;
\mbox{E}_{n-1}
\;\;\arrowschem{\mbox{\small U}}{}\;\;
\mbox{E}_n
\]
being driven forward by an enzyme U, with the different subscripts
indicating an additional degree of phosphorylation, and with constitutive
dephosphorylation.  
We will be interested in positive feedback from E$_n$ to U.

A typical way to model such a reaction is as follows.
We introduce variables $x_i(t)$, $i=1,\ldots ,n$ to indicate the fractional
concentrations of the various forms of the enzyme E 
(so that $x_1+\ldots +x_n\equiv 1$, and $x_i\geq 0$, for the solutions of physical
interest), and $u(t)%
\geq 0$ 
to indicate the concentration of U.
The differential equations are then as follows:
\beqn
\dot x_1 &=& 
-\sigma _1(u)\alpha _1(x_1) + \beta _2(x_2)
\\
\dot x_2 &=& 
\sigma _1(u)\alpha _1(x_1) - \beta _2(x_2)
-\sigma _2(u)\alpha _2(x_2) + \beta _3(x_3)
\\
\vdots
\\
\dot x_{n-1} &=& 
\sigma _{n-2}(u)\alpha _{n-2}(x_{n-2}) - \beta _{n-1}(x_{n-1})
-\sigma _{n-1}(u)\alpha _{n-1}(x_{n-1}) + \beta _{n}(x_{n})
\\
\dot x_{n} &=& 
\sigma _{n-1}(u)\alpha _{n-1}(x_{n-1}) - \beta _{n}(x_{n})
\,.
\eeqn
We make the assumptions that $\alpha _i$ and $\beta _i$ (respectively, $\sigma _i$)  are
differentiable functions $[0,\infty )\rightarrow [0,\infty )$ with positive (respectively, 
either positive or identically zero)
derivatives, and $\alpha _i(0)=\beta _i(0)=0$ 
and $\sigma _i(0)>0$
for each $i$.
(We allow some of the $\sigma _i$ to be constant, and in this manner represent
steps that are not controlled by U.)
Since we are interested in studying the effect of feeding back E$_n$,
we pick $y=x_n$.

Let us first prove that the characteristic is well-defined.
As we said, we are only interested in the solutions that lie in the
intersection $X$ of the plane $x_1+\ldots +x_n\equiv 1$ and the nonnegative orthant
in $\R^n$.  This set is easily seen to be invariant for the dynamics, and it
is convex, so the Brower fixed point theorem guarantees the
existence of an equilibrium in $X$, for any constant input $u(t)\equiv a$.
We next prove that this steady-state is unique.
Redefining if necessary the functions $\alpha _i$, we will assume without loss of
generality that $\sigma _i(a)=1$ for all $i$.
Let us introduce the 
nondecreasing
functions
\[
F_k \;=\; 
\beta _{k}^{-1}\circ \alpha _{k-1}\circ \beta _{k-1}^{-1}\circ \ldots \circ \beta _2^{-1}\circ \alpha _1
\]
for each $k=2,\ldots ,n$
and $F(r):=r+F_2(r)+\ldots +F_n(r)$.
This function is defined on some maximal interval $[0,M]$, consisting of those
$r$ such that $\alpha _1(r)$ belongs to the range of $\beta _2$,
$\alpha _2(\beta _2^{-1}(\alpha _1(r)))$ belongs to the range of $\beta _3$,
and so forth, and it is strictly increasing.
Moreover, for each equilibrium $x=(x_1,\ldots ,x_n)$, it holds that
$x_k=F_k(x_1)$, and therefore,
recalling that $x_1+\ldots +x_n=1$, $F(x_1)=1$.
Thus, if $x$ and $\widetilde x$ are two steady states, we have
$F(x_1)=F(\widetilde x_1)$.
Since $F$ is strictly increasing, it follows that $x_1=\widetilde x_1$, and therefore
that $x_k=F_k(x_1)=F_k(\widetilde x_1)=\widetilde x_k$ for all $k$, so uniqueness is shown.

We must prove stability.  For that, we first perform a change of coordinates:
\[
z_1=x_1,\;
z_2=x_1+x_2,\;
\ldots ,\;
z_{n-1}=x_1+\ldots +x_{n-1},\;
z_{n}=x_1+\ldots +x_n
\]
so that the equations in these new variables become (using that
$\dot z_k = (d/dt)(x_1+\ldots +x_k)$ and $x_k = z_k-z_{k-1}$ for $k>1$):
\beqn
\dot z_1 &=& -\sigma _1(u)\alpha _1(z_1) + \beta _2(z_2-z_1)\\
\vdots
\\
\dot z_{k} &=& 
-\sigma _{k}(u)\alpha _{k}(z_{k}-z_{k-1}) + \beta _{k+1}(z_{k+1}-z_{k})\\
\vdots
\\
\dot z_{n-1} &=& 
-\sigma _{n-1}(u)\alpha _{n-1}(z_{n-1}-z_{n-2}) + \beta _{n}(1-z_{n-1})
\eeqn
(and $z_n\equiv 1$).
When the input $u(t)$ is equal to any given constant,
the system described by the first $n-1$ differential equations, seen as
evolving in the subset of $\R^{n-1}$ where $0\leq z_1\leq z_2\leq \ldots \leq z_{n-1}\leq 1$,
is a {\em tridiagonal 
strongly
cooperative system}, and thus a theorem due to
Smillie (see~\cite{Sm}) insures that all trajectories converge to the set of
equilibria.
(The proof given in~\cite{smith-tridiagonal} is also valid when the
state-space is closed, as here.)
Moreover, linearizing at the equilibrium preserves the structure, so applying
the same result to the linearized system we know that we have in fact an
exponentially stable equilibrium.
Thus, characteristics are well defined.

It is easy to verify from our graph conditions that the system (in the new
coordinates) is 
monotone,
since
$\fract{df_i}{dz_j}>0$ for all
pairs $i\not= j$,
$\fract{df_i}{du}\leq 0$ 
for all $i$,
and
$\fract{dh}{dz_i}=0$ for all $i<n-1$ and 
$\fract{dh}{dz_{n-1}}<0$ (the output is $y=x_n=1-z_{n-1}$).
Excitability and transparency need not hold at boundary points; however,
Theorem~\ref{secondresult} still applies, because the closed-loop system is
strongly monotone.  To see this, it is enough to show that every trajectory
lies in the interior of $X$ for all $t>0$, since in the interior, the Jacobian
matrices are irreducible.  As the interior of $X$ is itself forward invariant
(see e.g.~\cite{monotone1}), it is sufficient to prove: for any $T>0$,
if $F$ is the set of $t\in [0,T]$ such that $x(t)$ is in the boundary of $X$
(relative to the linear space $x_1+\ldots +x_n=1$), then $F\not= [0,T]$.
Assume otherwise.
For each $i$, consider the closed set
$F_i=\{t\in [0,T] \st x_i(t)=0\}$, and note that
$\bigcup _iF_i=F$.
If $F_i$ would be nowhere dense for every $i$, then their union $F$ would be
nowhere dense, contradicting $F=[0,T]$.
Thus there is some $i$ so that $F_i$ contains an open interval 
$(a,b)\subseteq [0,T]$.
It follows that, for this $i$, $\dot x_i\equiv x_i\equiv 0$ on $(a,b)$, and
(looking at the equations) this implies that $x_{i\pm1}\equiv 0$ and, recursively,
we obtain $x_j\equiv 0$ for all $j$, contradicting $x_1+\ldots +x_n=1$.

\begin{figure}[htl]
\begin{center}
\includegraphics[width=4cm]{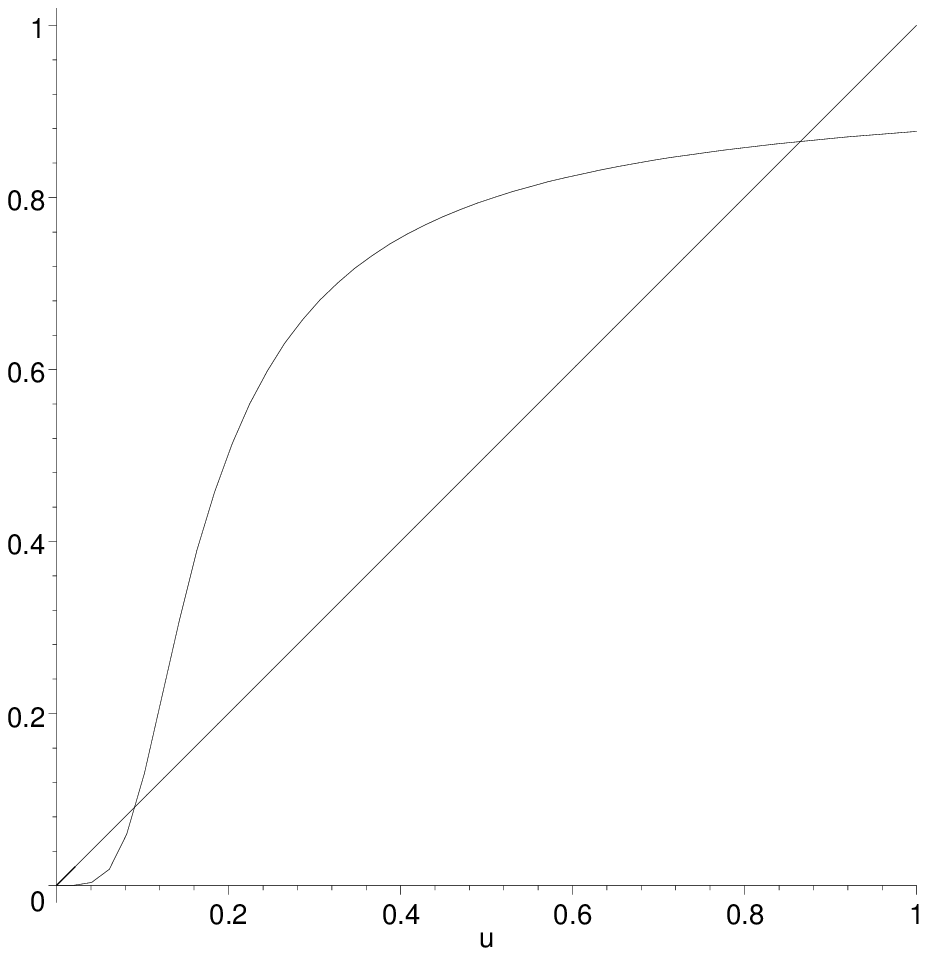}
\hspace{40pt}
\includegraphics[width=6cm]{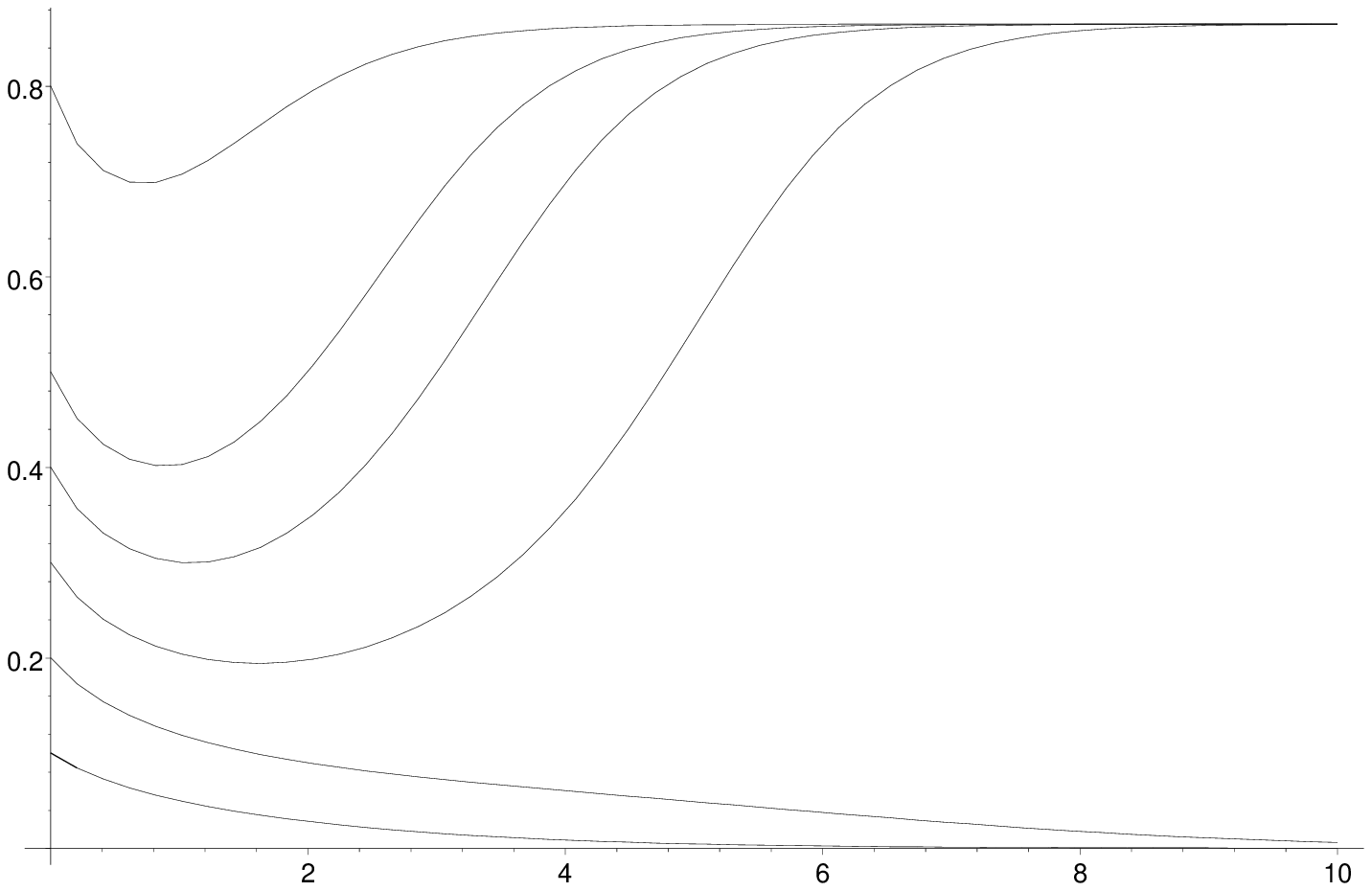}
\end{center}
\caption{Enzyme example: (a) Characteristic and (b) Simulations}
\label{7enzyme}
\end{figure}

As a numerical example, let us pick 
$\sigma _i(r)=(0.01+r)/(1+r)$,
$\alpha _i(r) = 10\,r/(1+r)$, and $\beta _i(r)=r/(1+r)$ for all $i$, and $n=7$.
(The constants have no biological significance, but the functional forms are
standard models of saturation kinetics.)
A plot of the characteristic is shown in Fig.~\ref{7enzyme}(a).
Since the intersection with the diagonal has three points as shown, we know
that {\em the closed-loop system (with $u=x_n)$ will have two stable and one
unstable equilibrium, and almost all trajectories converge to one of these two
stable equilibria.} 
To illustrate this convergence, we simulated six initial conditions,
in each case with $x_2(0)=\ldots =x_6(0)=0$ and with the following choices of
$x_7(0)$: 0.1, 0.2, 0.3, 0.4, 0.5, and 0.8 (and $x_1(0)=1-x_7(0)$).
A plot of $x_7(t)$ for each of these initial conditions is shown in
Fig.~\ref{7enzyme}(b); note the convergence to the two predicted steady states.

\section{External Stimuli, Thresholds and Hysteresis}

\begin{figure}[htl]
\begin{center}
\setlength{\unitlength}{1500sp}%
\begin{picture}(4824,2424)(2989,-5773)
\put(4201,-5161){\framebox(2400,1800){}}
\put(6601,-3961){\vector( 1, 0){1200}}
\put(3001,-3961){\vector( 1, 0){1200}}
\put(6601,-4561){\line( 1, 0){600}}
\put(7201,-4561){\line( 0,-1){1200}}
\put(7201,-5761){\line(-1, 0){3600}}
\put(3601,-5761){\line( 0, 1){1200}}
\put(3601,-4561){\vector( 1, 0){600}}
\put(3301,-3661){$v$}%
\put(7201,-3661){$w$}%
\put(7501,-5161){$y$}%
\put(3101,-5161){$u$}%
\end{picture}
\end{center}
\caption{Block diagram of unitary feedback system with external inputs}
\label{unfeedback}
\end{figure}
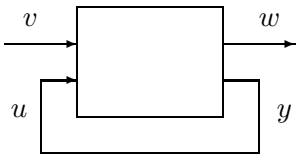

Throughout this section we investigate the behavior of positive
feedback interconnections of monotone systems which are in turn excited by some
exogenous input. In particular we consider interconnections of the
following type:
\be{withinputs}
\begin{array}{rcl}
\dot {x} &=& f ( x , u , v) \\
y & = & h_y (x) \\
w & = & h_w (x) 
\end{array}
\ee
along with the unitary feedback interconnection $u=y$.
The block diagram of such systems is shown in Fig.~\ref{unfeedback}.

We assume $f : X \times U \times V \rightarrow \R^n$ to be a 
locally Lipschitz function and that the system
(\ref{withinputs}) is a monotone control system with input $[u,v]$ and
output $[y,w]$ with respect to some ordering $\succeq_x$ of the
state-space $X$ and cross-product orders as far as inputs $[u,v]$ and 
outputs $[y,w]$ are concerned, (i.e.\ $[u_1 , v_1] \succeq_I [u_2,v_2]$
iff $u_1 \succeq_u u_2$ and $v_1 \succeq_v v_2$, $[y_1,w_1] \succeq_O
[y_2,w_2]$ iff $y_1 \succeq_y y_2$ and $w_1 \succeq_w w_2$).

 For each fixed value of the input $v$, systems as in
 (\ref{withinputs}) can be studied according to the techniques
 described previously.

\begin{figure}[htl]
\begin{center}
\setlength{\unitlength}{2000sp}%
\begin{picture}(6024,1974)(4189,-5173)
\thicklines
{\put(6601,-4561){\framebox(2400,1200){}}
}%
{\put(9001,-3961){\vector( 1, 0){1200}}
}%
{\put(6001,-3961){\vector( 1, 0){600}}
}%
{\put(5026,-4261){\framebox(975,600){}}
}%
{\put(4201,-3811){\vector( 1, 0){825}}
}%
{\put(9376,-3961){\line( 0,-1){1200}}
\put(9376,-5161){\line(-1, 0){4800}}
\put(4576,-5161){\line( 0, 1){1050}}
\put(4576,-4111){\vector( 1, 0){450}}
}%
{\put(4726,-4711){\dashbox{60}(4800,1500){}}
}%
\put(6151,-3811){$d$}%
\put(9826,-3811){$y=w$}%
\put(5120,-4036){$g(\cdot ,\cdot )$}%
\put(4000,-3736){$v$}%
\put(4276,-4336){$u$}%
\end{picture}

\end{center}
\caption{A special feedback configuration of SISO systems}
\label{unfeed2}
\end{figure}
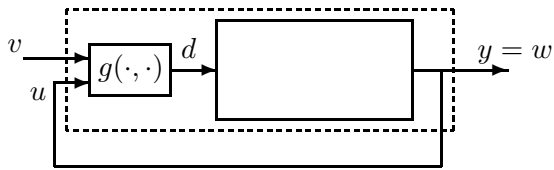

A special instance of systems of this kind is given by single-input,
single-output systems of
the following form:
\be{sisoposfeed}
\begin{array}{rcl}
\dot {x} &=& f(x,d) \\
      d &=& g(v,y) \\
      y &=& h(x)
\end{array}
\ee
where $g: V \times U \rightarrow \R$ is a monotone and locally Lipschitz function
(for instance $u,v \in \R_{\geq 0}$ and $g(v,y) = vy$ or $g(v,y) =v+y$). 
This structure (see Fig.~\ref{unfeed2}) is of
interest because it arises commonly in biological applications and is
particularly suited for a graphical analysis.

Next, we discuss the behavior of
 such interconnections in the presence of external stimuli.
In particular, in the case of multistable systems we prove the
existence of \emph{threshold} values of inputs which trigger
the transition among different equilibria.
 
\begin{figure}[htl]
\begin{center}
\includegraphics[width=6cm]{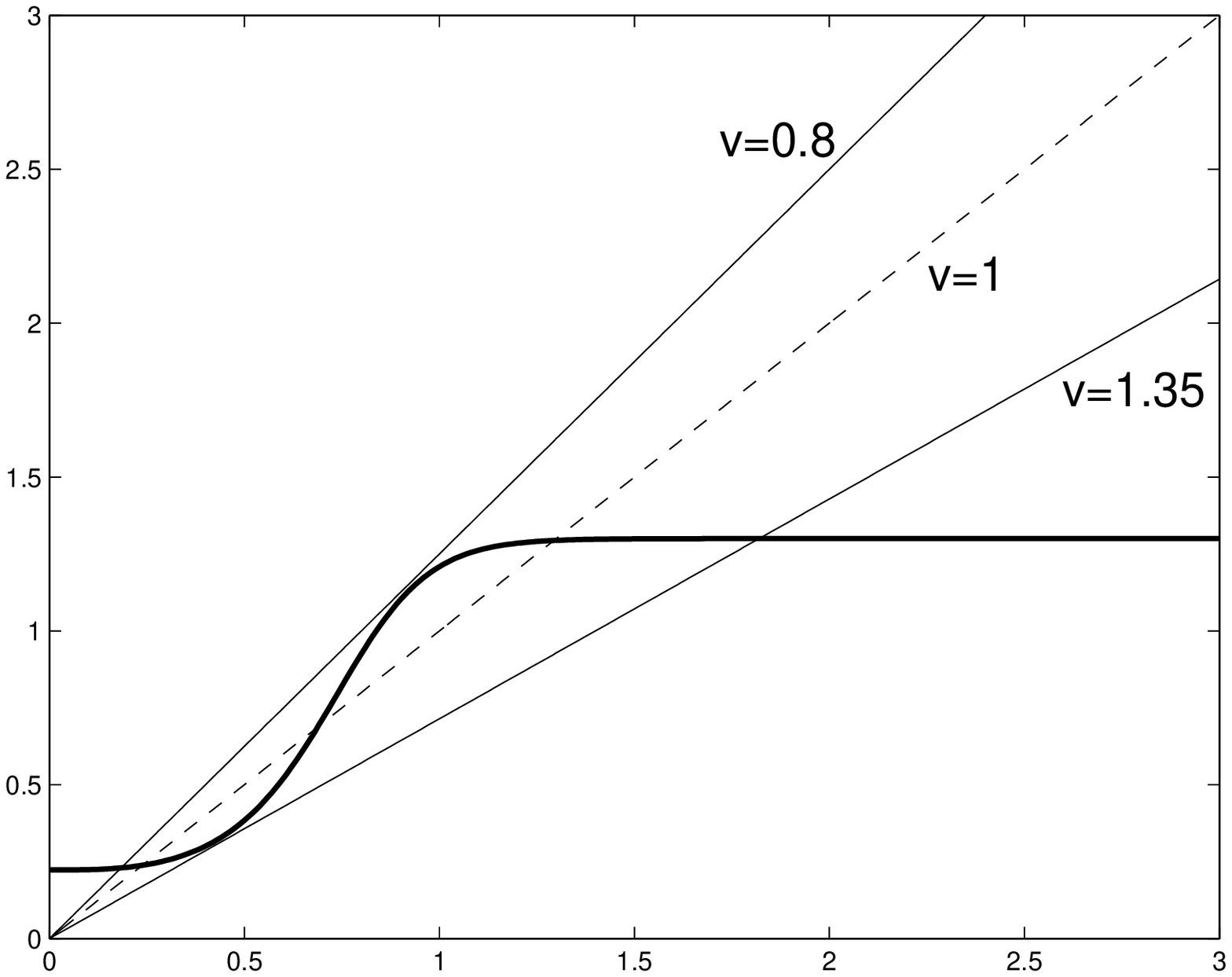}
\ \ \ 
\includegraphics[width=6cm]{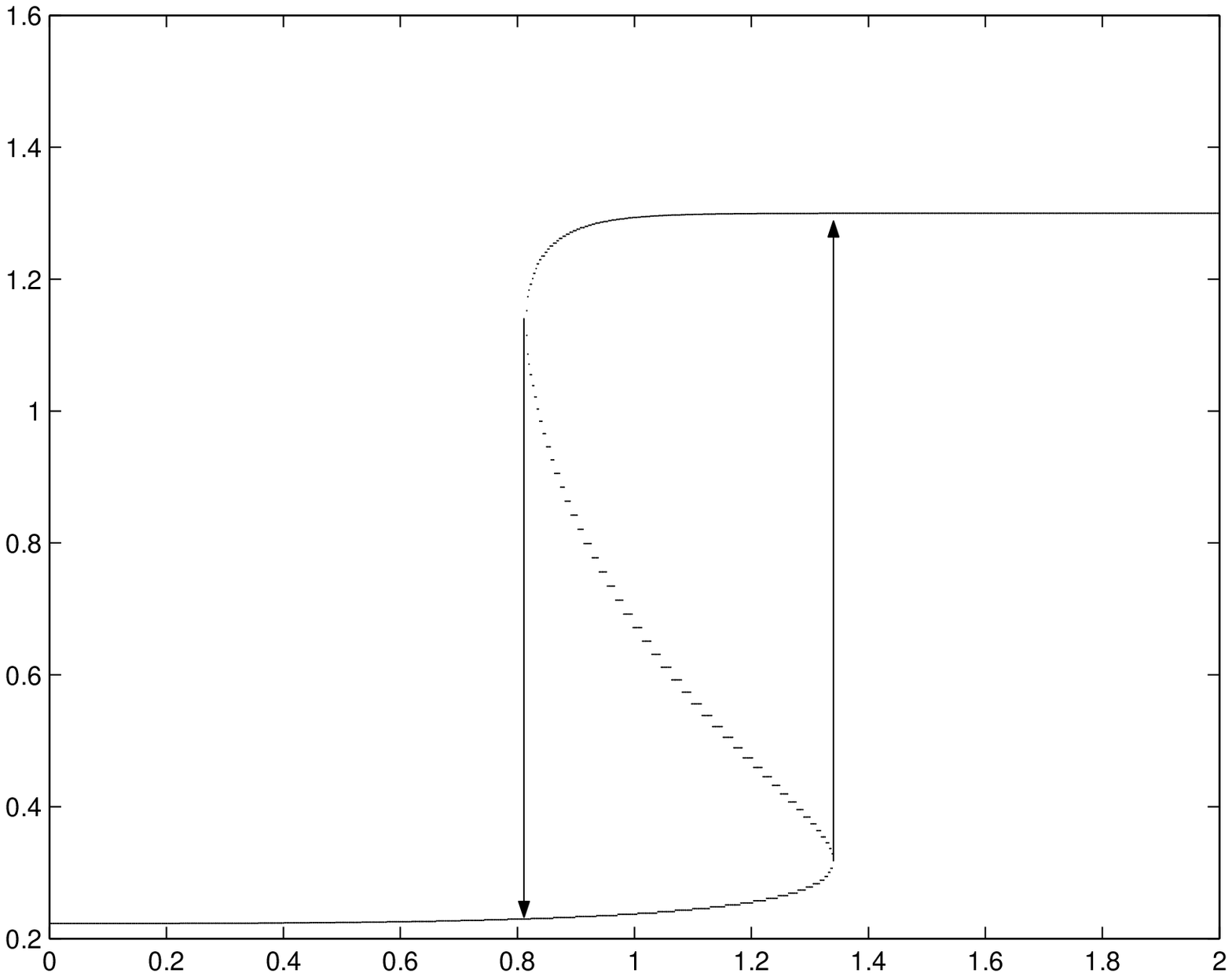}
\end{center}
\caption{Thresholds and hysteresis; horizontal axis is $u$
(resp. $v$) and vertical axis is $x_2$.}
\label{hist1}
\end{figure}

The above considerations suggest the possibility of studying 
 interconnections as in (\ref{withinputs}) by taking into account a
 parametrized family of I/O static characteristics in the $(u,y)$ plane, where the
 parameter is the exogenous input $v$. This type of analysis is
 very general and bifurcations can be traced by looking at the
 intersections of the parametrized I/O characteristic with the
 diagonal $u=y$. For the special structure (\ref{unfeed2}) instead,
 the study can be carried out in the $(d,y)$-plane allowing some
 intuitive simplifications.
 A single I/O characteristic is needed in fact, from $d$ to $y$, and
 equilibria correspond to intersections with the ``parametrized''
 family of functions $d = g(v,y)$, which also takes values in the
 $(d,y)$ plane.
 Although the analysis which follows is essentially a consequence
 of Theorem \ref{secondresult}, it is still worth pursuing, because it
 provides a solid theoretical justification to phenomena which are
 well described and understood in many biological applications.
 Consider again the system (\ref{ecoli2}), subject to the feedback
 interconnection $u=v \cdot y$. This results in the following
 set of equations:
 \be{ecoli3}
 \begin{array}{rcl}
 \dot {x}_1 &=& \frac{\alpha _1}{1+ (v \cdot x_2)^\beta } - x_1 \\ 
 \dot {x}_2 &=& \frac{\alpha _2}{1+ x_1^\gamma } - x_2 \\
 y & = & x_2. 
 \end{array}
 \ee
 We may therefore analyze the system by looking at the I/O
 static characteristic from $u$ to $y$, together with the $v$-parametrized
 family of lines $ y = u/v$.
 Fig.~\ref{hist1} illustrates a typical situation, corresponding
 here to the parameters value in the following table:
\begin{center}
\begin{tabular}{|c|c|c|c|}
\hline
$\gamma $ & $6$ & $\beta $ & $3$ \\
$\alpha _1$ & $1.3$ & $ \alpha _2$ & $1.3$ \\
\hline
\end{tabular}
\end{center}
Notice that for $v=1$  bistability is obtained; in particular two
equilibria are asymptotically stable and one is an unstable saddle
whose stable manifold behaves as a separatrix for the basins of
attractions of the stable equilibria.
Bifurcations occur at two different values of $v$, approximately
$v_1 \approx 0.8$ and $v_2 \approx 1.35$. This values correspond to 
the slopes of the tangent lines to the I/O characteristic. 
For all $v>v_2$ in fact there only exists one equilibrium, usually
referred to as the \emph{activated} equilibrium. For $v<v_1$ again only one
equilibrium occurs but corresponding to a \emph{non-activated} state. These values
play therefore the role of input \emph{thresholds} that may trigger
transition from the non-activated state to an activated one and vice
versa.
After a signal of amplitude bigger than $v_2$ is applied for a
sufficiently long time, the state will be in proximity of the
activated equilibria. Then, this level of output will be maintained
even after $v(t)$ drops below $v_2$, provided that $v_1 < v(t)$.
Further decrease of the $v(t)$ below $v_1$, for a sufficiently long
time,  will instead trigger
transition to a deactivated state, which is afterward maintained also
for higher values of $v(t)$, provided that $v(t)<v_2$.
This kind of behavior, known as hysteresis, has been observed in many
biological systems (see for instance \cite{ferr}, \cite{pomerening}).
In an actual experimental situation, one would block the feedback of $x_2$,
replacing the effect of $x_2$ by an experimentally set value of the input, and
the bifurcation diagram would be obtained directly from the I/O characteristic,
itself measured experimentally.  See~\cite{angeli-ferrell} for more discussion
of this topic.

\section{Why is Monotonicity Imposed ?}
\begin{figure}[htl]
\centerline{
\includegraphics[width=7cm]{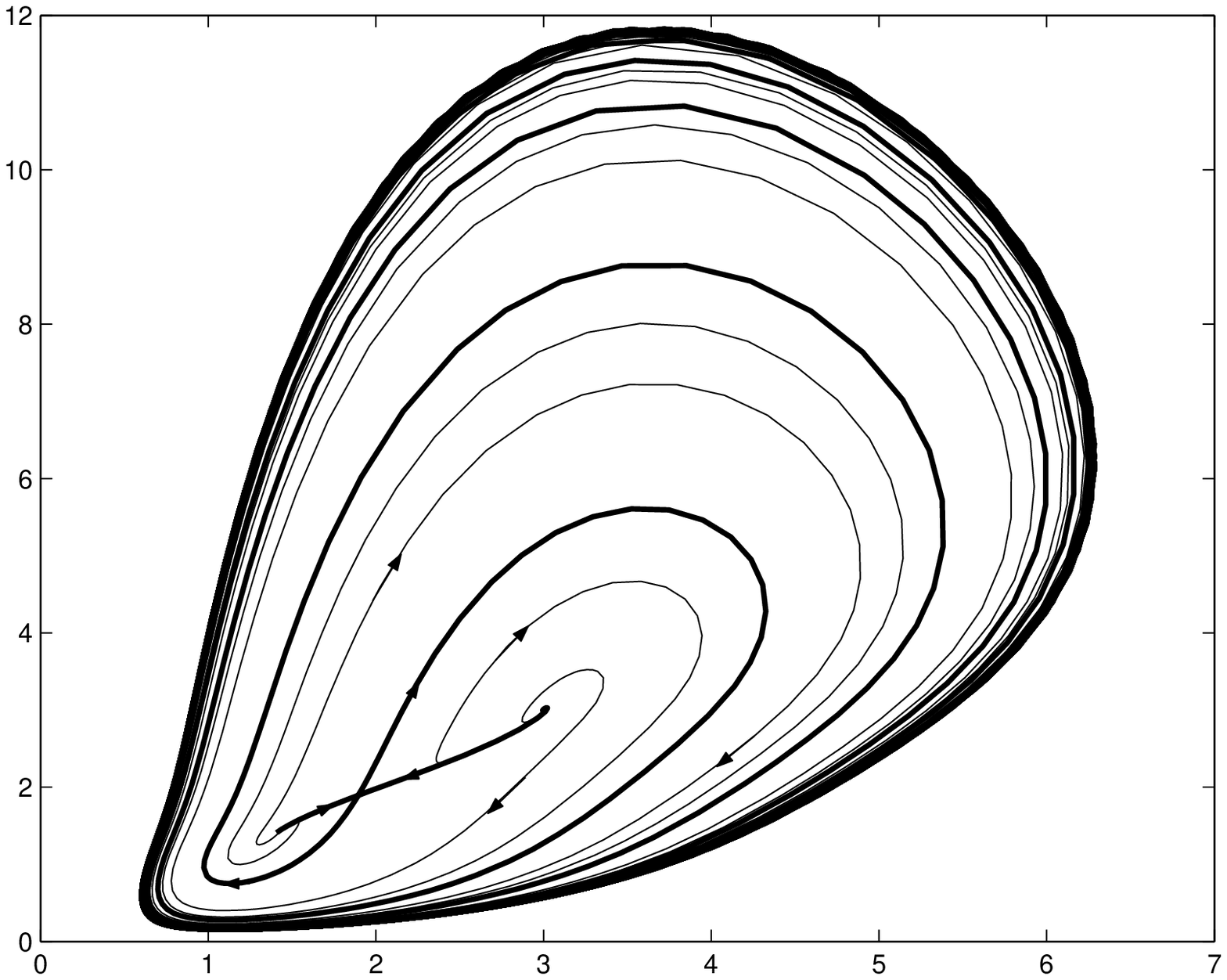}
\includegraphics[width=7cm]{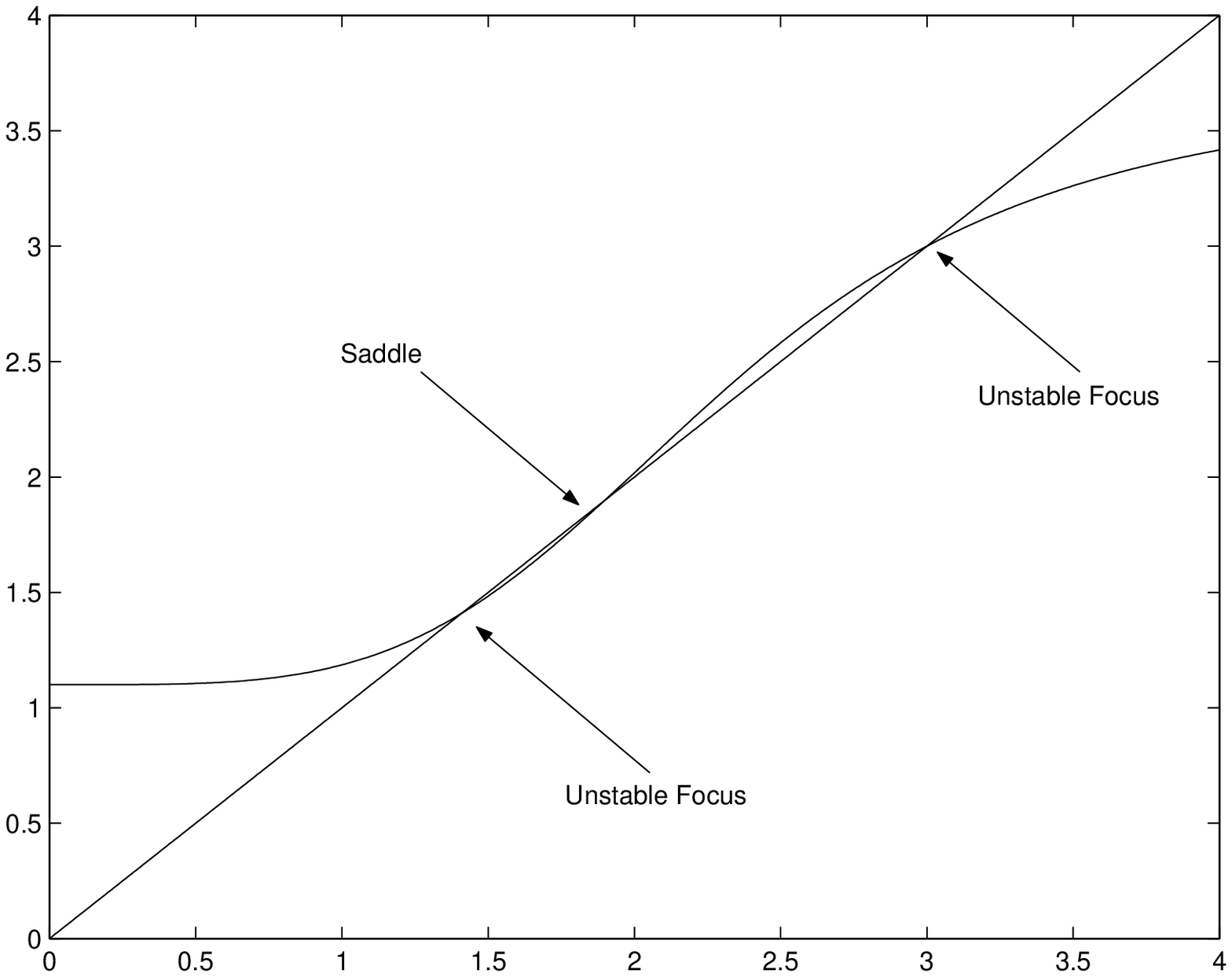}}
\caption{A stable limit cycle arising in a non-monotone feedback loop
  and the I/O static characteristic;
horizontal axis is $x_1$ (resp. $u$) and vertical axis is $x_2$ (resp. $y$)}
\label{limitcycle}
\end{figure}

Local analysis techniques based on the study of intersections of
static characteristics of interconnected systems or, in the two-dimensional case
of nullclines, are very common in mathematical biology. Our discussion
shows that for the class of monotone systems, under relatively mild
assumptions,  almost global convergence results can be obtained and
the investigation of the stability property of equilibria can be
carried out just by graphical inspection at the intersection points
of the I/O characteristics of systems in feedback.
In this section we show by means of an example how monotonicity
is a crucial assumption in this respect.
The following planar system (a predator-prey system):
\be{controesempio}
\begin{array}{rcl}
\dot {x}_1 &=& x_1 ( - x_1 + x_2) \\
\dot {x}_2 &=& 3 x_2 ( -x_1 + u ) \\
y &= & c + b \frac{x_2^4}{k+x_2^4}
\end{array}
\ee
evolving in $\R_{\geq 0}^2$, it is not monotone. However, it has a
well defined (monotonically increasing) I/O static characteristic
(see Fig.~\ref{limitcycle})
provided that $c, b, k \in \R_{>0}$. Moreover, for certain parameters
values, the I/O characteristics has $3$ (non-degenerate) fixed points.
 The closed-loop system resulting from the interconnection $u=y$,
 however, need not be globally converging at the
 set of equilibria. The simulations in Fig.~\ref{limitcycle} refer to the following values:
$c=1.1, b=361/140, k=405/14$.
Notice that the $3$ equilibria correspond to $2$ unstable foci and one
saddle point.

\section{Conclusions}

We have presented a general method for detecting multistability in a class of
positive feedback systems.  Our results apply when the original system has
certain properties (well-defined characteristic, monotonicity).  The results
can be used in conjunction with other techniques being developed, such as the
study of small-gain theorems for negative feedback interconnections
(cf.~\cite{monotone1}), in order to attempt to understand the behavior of
complex biological signaling interconnections by first breaking up the system
into smaller parts and then reconstituting the behavior of the entire system.

\appendix 
\section{Graphical characterizations of Transparency and Excitability}

\bl{eds-lemma}
Consider a scalar differential equation
$\dot x=f(x,u)$, evolving on a open subset of $\R$, 
where $u=(u_1,u_2,\ldots ,u_k)$ is a vector of input functions $u_j$ taking values
in nonempty sets $U_j\subseteq \R$.
We assume that $f$ is $C^1$ and that solutions are defined for all initial
states and all $t>0$, for any locally bounded inputs.
Suppose that the system is cooperative, that is, $f(x,u)$ is
nondecreasing as a function of $u_j$, for all $j$ 
(meaning that, for every $x$, $f(x,u)\geq f(x,v)$ if $u_j\geq v_j$ for all $j$).
Define the following set of indices:
\[
I_+ := \left\{j^\star\in \{1,\ldots ,k\} \st f(x,u) 
\mbox{\ is strictly increasing as a function of\ } u_{j^\star} \right\}
\]
(the strict increase condition meaning that $f(x,u)>f(x,v)$ for all $x,u,v$
such that $u_j\geq v_j$ for all $j$ and $u_{j^\star}>v_{j^\star}$),
and, for each $j\in \{1,\ldots ,k\}$, each $\varepsilon >0$, and each pair of inputs $u(\cdot )$
and $v(\cdot )$, the set of times: 
\[
S_{j,\varepsilon ,u,v} := \left\{t\in [0,\varepsilon ]\st u_{j}(t)>v_{j}(t)\right\}
\]
(possibly empty).
Pick any two inputs $u(\cdot )$ and $v(\cdot )$ such that $u\succeq v$ 
(i.e, $u_j(t)\geq v_j(t)$ for all $t$ and all $j$) and
suppose that either:
\ben
\item
there is some $j^\star\in I_+$ is such that the Lebesgue measure
$\mu \left(S_{j^\star,\varepsilon ,u,v}\right)>0$ for each $\varepsilon >0$, or
\item
$u\succ v$ and $I_+=\{1,\ldots ,k\}$.
\een
Then, for each initial state $\xi $, the respective solutions for these two
inputs satisfy
$x(t,\xi ,u)>x(t,\xi ,v)$ for all $t>0$.
\els

\bpr
Take two such inputs and initial state.  Since the system is monotone, we know
that $x(t,\xi ,u)\geq x(t,\xi ,v)$ for all $t>0$, but we need to prove that the
strict inequality holds for all $t$.
So suppose that there is some $T>0$ such that $x(T,\xi ,u)=x(T,\xi ,v)=\zeta $.

We claim that, then, $x(s,\xi ,u)=x(s,\xi ,v)$ for all $s<T$.
This fact is an easy consequence of comparison arguments based upon
monotonicity (see~\cite{rouche}).  We provide a proof here for the reader's
convenience.
Suppose that $\xi '=x(S,\xi ,u)>x(S,\xi ,v)=\xi ''$ for some $S<T$, and
consider the following system of two differential equations:
\beqn
\dot x&=&f(x,u)\\
\dot z&=&f(z,v) \,.
\eeqn
Pick a sequence $(a_k,b_k)\rightarrow (\zeta ,\zeta )$ with the property that $a_k<b_k$ for all
$k$ (for instance, $b_k\equiv \zeta $ and $a_k=\zeta -1/k$ for $k$ large enough).
Let $\Phi $ be the map that sends initial states $(x(S),z(S))$ at time $S$ into
states $(x(T,x(S),u_S),z(T,z(S),v_S))$ at time $T$, where $u_S$ and $v_S$ are
the inputs restricted to times $\geq S$ (we may think of $\Phi $ as the time $T-S$
flow of the two-dimensional system).
As this map is a diffeomorphism, there exists a sequence
$(c_k,d_k)\rightarrow (\xi ',\xi '')$ such that $\Phi (c_k,d_k)=(a_k,b_k)$ for all $k$.
Since $\xi '>\xi ''$, it follows that $c_k>d_k$ for some $k$.
This means that the solution of the above system, with initial state
$x(S)=c_k>d_k=z(S)$ satisfies $x(T)<z(T)$, contradicting the monotonicity of
the original system.

Since $x(s)=x(s,\xi ,u)=x(s,\xi ,v)$ for all $s<T$, we may take derivatives with
respect to time to conclude that
\[
f(x(s),u(s)) = f(x(s),v(s))
\]
for all $s\in [0,T]$.  

Suppose that there is some $j^\star\in I_+$ so that
$\mu \left(S_{j^\star,T,u,v}\right)>0$.
Then we may pick a time $t\in S_T$ so that
$u_{j^\star}(t)>v_{j^\star}(t)$ and also $f(x(t),u(t)) = f(x(t),v(t))$.
This contradicts the strict increase assumption $j^\star\in I_+$.

Suppose instead that $u\succ v$.  Then, we claim, there is some
$j^\star$ such that $\mu \left(S_{j^\star,\varepsilon ,u,v}\right)>0$ for all $\varepsilon >0$.
Indeed, if this claim were false, then there would be for each $j$ some
$\varepsilon _j>0$ such that $\mu \left(S_{j,\varepsilon _j,u,v}\right)=0$, which implies that
also $\mu \left(S_{j,\varepsilon ,u,v}\right)=0$, where $\varepsilon =\min \varepsilon _j$.
Thus the union of these sets has measure zero, that is,
$u_j(t)=v_j(t)$ for all $j$ and all $t\in [0,\varepsilon ]$, contradicting $u\succ v$.
Since $j^\star\in \{1,\ldots ,m\}=I_+$, we have reduced to the first case.
\epr

For one-dimensional systems, Theorem~\ref{excitability2} can be strengthened
into a necessary and sufficient statement.  The proof of the Theorem will
recursively use this result.

\bc{scalarmonotone}
Let $\dot {x}=f(x,u)$ be a scalar cooperative system as in Lemma~\ref{eds-lemma},
and assume that this system has a well-defined incidence graph.
Then, the system is excitable if and only if
$I_+=\{1,\ldots ,k\}$, and it is weakly excitable if and only if
$I_+\not= \emptyset$
\ecs

\bpr
Suppose that $I_+=\{1,\ldots ,k\}$, and pick any two inputs $u\succ v$.
The second case in the Lemma then gives that 
$x(t,\xi ,u)>x(t,\xi ,v)$ for all $t>0$, and this proves excitability.
If, instead, $u\gg v$ and we know that $I_+\not= \emptyset$, we pick any
$j^\star\in I_+$ and use the fact that $u\gg v$ implies that 
$S_{j^\star,\varepsilon ,u,v}=[0,\varepsilon ]$ for all $\varepsilon $, so the first case in the Lemma then
gives that $x(t,\xi ,u)>x(t,\xi ,v)$ for all $t>0$, and this proves weak
excitability.

To prove the converse implications, we first consider the case  
$I_+=\emptyset$. 
By definition of incidence graph, this means that
$\frac{\partial f}{\partial u_j} (x,u) \equiv 0$ for all $j$.
Thus, solutions do not depend on input signals, and this contradicts weak
excitability.
If, instead, we only know that some $j^\star\not\in I_+$, then we have that
$\frac{\partial f}{\partial u_{j^\star}} (x,u) \equiv 0$, and we may take any initial state $\xi $, and
any two inputs $u(\cdot )$ and $v(\cdot )$ with the property that
$u_\ell(t)=v_\ell(t)$ for all $t$ and all $\ell\not= j^\star$,
and $u_{j^\star}(t)>v_{j^\star}(t)$; since
$x(\cdot ,\xi ,u)\equiv x(\cdot ,\xi ,v)$, we have that $u\succ v$ but it is false that
$x(t,\xi ,u)>x(t,\xi ,v)$ for $t>0$, contradicting excitability.
\epr

\subsubsection*{Proof of Theorem~\protect{\ref{excitability2}}}

By appropriate coordinate changes $x_i\mapsto (-1)^{\delta _i}x_i$, as done
in~\cite{monotone1}, one may restrict attention to cooperative systems.

Consider a cooperative system which admits an incidence graph and assume that
each vertex $x_i$ is reachable from some input vertex $u_j$. 
Let $\xi $ be an arbitrary initial condition and let
$u$, $v$ be arbitrary input signals satisfying $v\gg u$.
We know that $x(t,\xi ,v) \succeq x(t,\xi ,u)$ for all $t$, and must
show the strict inequality for all state components, i.e.,
$x(t,\xi ,v) \gg x(t,\xi ,u)$. 
We will prove this by induction, exploiting repeatedly Lemma~\ref{eds-lemma}. 
To this end, we decompose the system in sublayers, based on the following
notion of distance among vertices of a graph:
\be{distance}  
d(v \rightarrow w) = 
\min \{ L(\mathcal{P}) : \mathcal{P}_0=v \textrm{ and } \mathcal{P}_L = w \},
\ee   
i.e. $d(v \rightarrow w)$ denotes the shortest length among all paths which link $v$ to
$w$.
Furthermore, for each state vertex $x_i$ of the incidence graph, we define the
following integer:
\be{level}
D (x_i) = \inf_j \; d (u_j \rightarrow x_i)
\ee
(in words: $D(x_i)$ corresponds to the minimum distance from some input vertex
$u_j$  to the state vertex $x_i$).
Notice, by the reachability assumption, that $D(x_i)$ is well-defined ($<+\infty $)
for every $i \in \{ 1, \ldots n \}$. 
We say that $x_i$ belongs to the $k$-th sublayer, if $D (x_i)=k$. 

Consider any state coordinate $x_i$ so that $D(x_i) =1$ (such a coordinate
always exists).
We view $\dot {x}_i = f_i (x,u)$ as a scalar (cooperative) system, forced by the
inputs $u$ and $x_{k}$ for all $k \neq i$.
As $D(x_i)=1$, there exists $j^\star$ such that $f_i (x,u)$ is strictly
monotone as a function of $u_{j^\star}$.
Since $u\gg v$, $v_{j^\star} (t) > u_{j^\star} (t)$ for almost all $t \geq 0$.
Then the first part of Lemma~\ref{eds-lemma} allows us to conclude that
$x_i(t,\xi ,v) > x_i (t,\xi ,u)$ for all $t > 0$.
This shows that the strict inequality holds for all state components belonging
to the first sublayer.

Proceeding by induction, any component belonging to the $i$-th sublayer is
reachable in one step from at least some component $x_j$ belonging to 
the $(i-1)$st sublayer (strict monotonicity of $f_i (x,u)$ with respect to
$x_j$), and once again viewing $\dot {x}_i = f_i (x,u)$ as a scalar (cooperative)
system, this time using $x_j(t)$ as the input, we conclude that
$x_i(t,\xi ,v) > x_i (t,\xi ,u)$ for all $t >0$.
 
This completes the proof for the case of weak excitability. 
Next we consider the case of excitability.

Assume that $v \succ u$. 
Arguing as in the proof of Lemma~\ref{eds-lemma}, we know that there exists an
integer $j^\star$ so that 
$\mu \left(S_{j^\star,\varepsilon ,u,v}\right)>0$ for each $\varepsilon >0$.
We again prove the result by induction by considering a sublayer
decomposition, this time taken by looking at graph distances
with respect to this particular input vertex $j^\star$,
i.e.: $D(x_i):= d(u_{j^\star}  \rightarrow x_i )$.
By the reachability assumption in the case of weak excitability, 
$D(x_i)$ is well-defined for all $i\in \{1,2,\ldots n\}$.

Pick any state component $x_i$ for which $D(x_i) =1$.
Once again, we view $\dot {x}_i = f_i (x,u)$ as a scalar cooperative system,
forced by the inputs $u$ and $x_{k}$ for all $k \neq i$. 
In particular $f_i (x,u)$ is strictly monotone with respect
to $u_{j^\star}$, and we may apply the Lemma.
Arguing by induction, any component belonging to the $i$-th sublayer is reachable in one step by some component belonging to the
$(i-1)$st sublayer, and therefore a similar argument applies, yielding
$x_i(t,\xi ,v) > x_i (t,\xi ,u)$ for all $t >0$. 
\qed

\subsubsection*{Sketch of Proof of Theorem~\protect{\ref{transparency2}}}

Consider an arbitrary pair of ordered initial conditions $\xi _1 \succ \xi _2$. 
By monotonicity and uniqueness of solutions, we have 
$x(t,\xi _1,u) \succ x(t,\xi _2,u)$ for all $t \geq 0$.
Arguing as earlier, we know that there must exist some index $j^\star$ so that,
for all $\varepsilon >0$, the set
$\{ t\in [0,\varepsilon ] \st x_{j^\star} (t,\xi _1,u) > x_{j^\star}(t,\xi _2,u) \}$ has
non-zero measure.

We claim that $x_i (t,\xi _1,u) > x_i (t,\xi _2,u)$ for every vertex $x_i$ which
is reachable from the vertex $x_{j^\star}$, and denote with
$\mathcal{R}_{j^\star}$ the set of such $x_i$s.  The claim can be shown
inductively by an argument analogous to the one employed in the proof of
Theorem \ref{excitability2}.

By the graph reachability condition (either weak or strong), for all (some)
output vertices $y_j$ there exists at least one $x_i \in \mathcal{R}_{j^\star}$
so that $x_i \rightarrow y_j$ is an edge of the incidence graph. 
Thus, $h_j ( x(t,\xi _1,u) ) > h_j (x (t,\xi _2,u) )$ for all $t >0$ for
all such $j$s.
\qed

\section{Proof of Lemma~\protect{\ref{Perronfrob}}}

\bpr Consider the exponential map $\xi \rightarrow e^{At} \xi $. By positive
invariance of $K$, for each $t > 0$ the exponential is a linear map from $K$ to $K$.
Moreover, for $t$ sufficiently small $t$, $t\rightarrow e^\lambda t$ is one-to-one on the
spectrum of $A$. Thus, by Lemma A.3.3 in \cite{mct}, the geometric multiplicity of 
$e^{\lambda _i t}$ as an eigenvalue of the exponential map is the same
as that of $\lambda _i$ as an eigenvalue of $A$, with the same respective
eigenvectors. Therefore, we can study the 
spectrum of $A$ by looking at the spectrum of its exponential map for
$t$ sufficiently small.
By the Perron-Frobenius Theorem, there exists a real positive eigenvalue
$\mu $, with eigenvector $v \in K$, which is dominant in the sense
that $\mu = \rho (e^{At})$ (eigenvalue of maximum modulus).
Therefore, we conclude that $\lambda := \log( \mu )/t$ is an eigenvalue
for $A$, relative to the same eigenvector $v \in K$,  and $\textrm{Re}(\lambda ) \geq \textrm{Re} (\lambda _i)$ for
all $\lambda _i \in \textrm{Spec}(A)$.
\epr

\edo
restart:with(plots):with(DEtools):
K:=1:
k:=1:
a:=10:
b:=1:
c:=1:
# changed from zero to this
d:=0.01:
speedup:=1000:
f:= r\rightarrow a*r/(K+r):
g:= r\rightarrow b*r/(k+r):
h:= r\rightarrow d+c*r/(k+r):
lhs0:=1:
lhs1:=speedup*(-h(u(t))*f(x1(t)) + g(x2(t))):
lhs2:=speedup*(h(u(t))*f(x1(t)) - g(x2(t)) - h(u(t))*f(x2(t)) + g(x3(t))):
lhs3:=speedup*(h(u(t))*f(x2(t)) - g(x3(t)) - h(u(t))*f(x3(t)) + g(x4(t))):
lhs4:=speedup*(h(u(t))*f(x3(t)) - g(x4(t)) - h(u(t))*f(x4(t)) + g(x5(t))):
lhs5:=speedup*(h(u(t))*f(x4(t)) - g(x5(t)) - h(u(t))*f(x5(t)) + g(x6(t))):
lhs6:=speedup*(h(u(t))*f(x5(t)) - g(x6(t)) - h(u(t))*f(x6(t)) + g(x7(t))):
lhs7:=speedup*(h(u(t))*f(x6(t)) - g(x7(t))):
lhs1+lhs2+lhs3+lhs4+lhs5+lhs6+lhs7:
init:=u(0)=0,x1(0)=1,x2(0)=0,x3(0)=0,x4(0)=0,x5(0)=0,x6(0)=0,x7(0)=0:
vars:=u(t),x1(t),x2(t),x3(t),x4(t),x5(t),x6(t),x7(t):
sys:=diff(u(t),t)=lhs0,
     diff(x1(t),t)=lhs1,
     diff(x2(t),t)=lhs2,
     diff(x3(t),t)=lhs3,
     diff(x4(t),t)=lhs4,
     diff(x5(t),t)=lhs5,
     diff(x6(t),t)=lhs6,
     diff(x7(t),t)=lhs7:
sol:=dsolve({sys,init},{vars},type=numeric):

#p1:=odeplot(sol,[[t,u(t)]],0..1,thickness=3,color=black):
p2:=odeplot(sol,[[t,x7(t)]],0..1,thickness=3,color=black):
p3:=plot(u,u=0..1,thickness=3,color=black):
display(p3,p2);

# now solve closed-loop
lhs1:=-h(x7(t))*f(x1(t)) + g(x2(t)):
lhs2:=h(x7(t))*f(x1(t)) - g(x2(t)) - h(x7(t))*f(x2(t)) + g(x3(t)):
lhs3:=h(x7(t))*f(x2(t)) - g(x3(t)) - h(x7(t))*f(x3(t)) + g(x4(t)):
lhs4:=h(x7(t))*f(x3(t)) - g(x4(t)) - h(x7(t))*f(x4(t)) + g(x5(t)):
lhs5:=h(x7(t))*f(x4(t)) - g(x5(t)) - h(x7(t))*f(x5(t)) + g(x6(t)):
lhs6:=h(x7(t))*f(x5(t)) - g(x6(t)) - h(x7(t))*f(x6(t)) + g(x7(t)):
lhs7:=h(x7(t))*f(x6(t)) - g(x7(t)):
lhs1+lhs2+lhs3+lhs4+lhs5+lhs6+lhs7:
vars:=x1(t),x2(t),x3(t),x4(t),x5(t),x6(t),x7(t):
sys:=     diff(x1(t),t)=lhs1,
     diff(x2(t),t)=lhs2,
     diff(x3(t),t)=lhs3,
     diff(x4(t),t)=lhs4,
     diff(x5(t),t)=lhs5,
     diff(x6(t),t)=lhs6,
     diff(x7(t),t)=lhs7:

init1:=x1(0)=0.9,x2(0)=0,x3(0)=0,x4(0)=0,x5(0)=0,x6(0)=0,x7(0)=0.1:
sol1:=dsolve({sys,init1},{vars},type=numeric):
p1:=odeplot(sol1,[[t,x7(t)]],0..10,thickness=3,color=black):

init2:=x1(0)=0.8,x2(0)=0,x3(0)=0,x4(0)=0,x5(0)=0,x6(0)=0,x7(0)=0.2:
sol2:=dsolve({sys,init2},{vars},type=numeric):
p2:=odeplot(sol2,[[t,x7(t)]],0..10,thickness=3,color=black):

init3:=x1(0)=0.7,x2(0)=0,x3(0)=0,x4(0)=0,x5(0)=0,x6(0)=0,x7(0)=0.3:
sol3:=dsolve({sys,init3},{vars},type=numeric):
p3:=odeplot(sol3,[[t,x7(t)]],0..10,thickness=3,color=black):

init4:=x1(0)=0.6,x2(0)=0,x3(0)=0,x4(0)=0,x5(0)=0,x6(0)=0,x7(0)=0.4:
sol4:=dsolve({sys,init4},{vars},type=numeric):
p4:=odeplot(sol4,[[t,x7(t)]],0..10,thickness=3,color=black):

init5:=x1(0)=0.5,x2(0)=0,x3(0)=0,x4(0)=0,x5(0)=0,x6(0)=0,x7(0)=0.5:
sol5:=dsolve({sys,init5},{vars},type=numeric):
p5:=odeplot(sol5,[[t,x7(t)]],0..10,thickness=3,color=black):

init8:=x1(0)=0.2,x2(0)=0,x3(0)=0,x4(0)=0,x5(0)=0,x6(0)=0,x7(0)=0.8:
sol8:=dsolve({sys,init8},{vars},type=numeric):
p8:=odeplot(sol8,[[t,x7(t)]],0..10,thickness=3,color=black):

display(p1,p2,p3,p4,p5,p8);